\documentclass[preprint2]{aastex6}

\usepackage{xspace}

\usepackage[T1]{fontenc}
\usepackage{ifthen}

\newcommand{\Kepler}{\textit{Kepler}\xspace} 
\newcommand{\ktwo}{\textit{K2}\xspace}

\newcommand{\SME}{\texttt{SME}\xspace}

\newcommand{\Gaia}{\textit{Gaia}\xspace} 
\newcommand{\Mstar}{\ensuremath{M_{\star}}\xspace}
\newcommand{\Rstar}{\ensuremath{R_{\star}}\xspace} 
 
\newcommand{\fe}{\ensuremath{\mathrm{[Fe/H]}}\xspace}
\newcommand{\teff}{$T_{\mathrm{eff}}$\xspace}  
\newcommand{\logg}{\ensuremath{\log g}\xspace} 
\newcommand{\vsini}{\ensuremath{v \sin i}\xspace}

\newcommand{\Mp}{\ensuremath{M_{P}}\xspace} 
\newcommand{\Mcore}{\ensuremath{M_{\mathrm{core}}}\xspace}
\newcommand{\Menv}{\ensuremath{M_{\mathrm{env}}}\xspace}
\newcommand{\Mcmf}{\Mcore/\Mp}
\newcommand{\fenv}{\ensuremath{f_{\mathrm{env}}}\xspace}
\newcommand{\Mfenv}{\Menv/\Mp}
\newcommand{\Qprime}{\ensuremath{Q^{\prime}}\xspace}
\newcommand{\vesc}{\ensuremath{v_{esc}}\xspace}
\newcommand{\vorb}{\ensuremath{v_{orb}}\xspace}

\newcommand{\Rp}{\ensuremath{R_P}\xspace}
\newcommand{\Teq}{$T_{\mathrm{eq}}$\xspace}
\newcommand{\Sinc}{\ensuremath{S_{\mathrm{inc}}}\xspace}

\newcommand{\ms}{\ensuremath{\mathrm{m}\,\mathrm{s}^{-1}}\xspace}
\newcommand{\kms}{\ensuremath{\mathrm{km}\,\mathrm{s}^{-1}}\xspace}
\newcommand{\msyr}{\ensuremath{\mathrm{m}\,\mathrm{s}^{-1}\,\mathrm{yr}^{-1}}\xspace}
\newcommand{\AU}{\ensuremath{\mathrm{AU}}\xspace}
\newcommand{\Se}{\ensuremath{S_{\oplus}}\xspace}
\newcommand{\Me}{\ensuremath{M_{\oplus}}\xspace} 
\renewcommand{\Re}{\ensuremath{R_{\oplus}}\xspace} 

\newcommand{\gcc}{g~cm$^{-3}$\xspace}
\newcommand{\Rsun}{\ensuremath{R_{\odot}}\xspace }
\newcommand{\Msun}{\ensuremath{M_{\odot}}\xspace}

\newcommand{\bjdtdb}{\ensuremath{\mathrm{BJD}_\mathrm{TBD}}\xspace}
\def\deg{\ensuremath{^{\circ}}}
\newcommand{\rrat}{\ensuremath{\Rp/\Rstar}\xspace} 

\newcommand{\lonperi}{\ensuremath{\omega_{\star}}\xspace}

\newcommand{\sqrtecosw}{\ensuremath{\sqrt{e} \cos \lonperi}\xspace}
\newcommand{\sqrtesinw}{\ensuremath{\sqrt{e} \sin \lonperi}\xspace}
\newcommand{\dvdt}{\ensuremath{\dot{\gamma}}\xspace}

\newcommand{\sigjit}[1]{
        \ensuremath{
                \ifthenelse{\equal{#1}{}}{\sigma_\mathrm{jit}}{\sigma_\mathrm{jit,#1}}}
        \xspace
}
\newcommand{\gam}[1]{\ensuremath{\gamma_\mathrm{#1}}\xspace}
\newcommand{\K}[1]{\ensuremath{K_\mathrm{#1}}\xspace}
\newcommand{\e}[1]{\ensuremath{e_\mathrm{#1}}\xspace}

\newcommand{\dbic}{\ensuremath{\Delta\mathrm{BIC}}\xspace}
\usepackage{xstring} 

\newcommand{\ktwotwentysevenstar}[1]{%
  \IfEqCase{#1}{%
    {teff}{ 5248 \pm 60}
    {logg}{ 4.48 \pm 0.05}%
    {fe}{ 0.13 \pm 0.04}%
    {vsini}{2.3}%
    {mass}{ 0.866_{ -0.023 }^{+0.029} }
    {radius}{ 0.885 \pm 0.043  }%
    {agegyr}{ 10.3_{ -5.2 }^{+3.3} }%
    {vmag}{ 12.64 \pm 0.02 }
  }[\PackageError{tree}{Undefined option to tree: #1}{}]%
}%

\newcommand{\ktwotwentysevencirc}[1]{%
  \IfEqCase{#1}{%
  {gamma_harps-n}{ 6.4 \pm 2.2  }%
  {gamma_hires}{ -2.7 \pm 2.0  }%
  {k1}{ 9.9 \pm 2.0  }%
  {jit_hires}{ 6.1_{ -1.4 }^{+2.0} }%
  {gamma_harps}{ -0.3 \pm 3.8  }%
  {jit_harps-n}{ 4.7 \pm 2.4  }%
  {jit_harps}{ 6.9_{ -3.2 }^{+5.9} }%
  {lnprobability}{ -136.6_{ -3.1 }^{+2.1} }%
  {e1}{  }%
  {P1}{ 6.771315 \pm 0.000085  }%
{T01}{ 1979.84484 \pm 0.00057  }%
{RpRstar1}{ 0.0474 \pm 0.0010  }%
{Rp1}{ 4.48 \pm 0.23  }%
{Mpsini1}{ 26.7 \pm 5.3  }%
{rhop1}{ 1.61 \pm 0.42  }%
{Teq1}{ 902 \pm 28  }%
{Lstar1}{ 0.520_{ -0.055 }^{+0.071} }%
{a1}{ 0.06702 \pm 0.00071  }%
{Sinc1}{ 116_{ -13 }^{+16} }%
  }[\PackageError{tree}{Undefined option to tree: #1}{}]%
}%

\newcommand{\ktwotwentysevenecc}[1]{%
  \IfEqCase{#1}{%
{secosw1}{ 0.34_{ -0.18 }^{+0.11} }%
{gamma_harps-n}{ 6.4 \pm 2.2  }%
{gamma_hires}{ -2.2 \pm 1.6  }%
{k1}{ 11.8 \pm 1.8  }%
{jit_hires}{ 4.4_{ -1.2 }^{+1.8} }%
{gamma_harps}{ 0.2 \pm 2.7  }%
{jit_harps-n}{ 4.8_{ -2.1 }^{+2.6} }%
{jit_harps}{ 4.1_{ -2.6 }^{+4.8} }%
{sesinw1}{ 0.33_{ -0.27 }^{+0.18} }%
{lnprobability}{ -130.4_{ -3.6 }^{+2.6} }%
{e1}{ 0.251 \pm 0.088  }%
{P1}{ 6.771315 \pm 0.000085  }%
{T01}{ 1979.84484 \pm 0.00057  }%
{RpRstar1}{ 0.0474 \pm 0.0010  }%
{Rp1}{ 4.48 \pm 0.23  }%
{Mpsini1}{ 30.9 \pm 4.6  }%
{rhop1}{ 1.87 \pm 0.41  }%
{Teq1}{ 902 \pm 28  }%
{Lstar1}{ 0.519_{ -0.055 }^{+0.072} }%
{a1}{ 0.06702 \pm 0.00071  }%
{Sinc1}{ 116_{ -12 }^{+16} }%
}[\PackageError{tree}{Undefined option to tree: #1}{}]%
}%

\newcommand{\ktwothirtytwostar}[1]{%
  \IfEqCase{#1}{%
    {teff}{ 5275 \pm 60}
    {logg}{ 4.49 \pm 0.05}%
    {fe}{ -0.02 \pm 0.04}%
    {vsini}{ 0.7 }%
    {mass}{ 0.856 \pm 0.028  }%
    {radius}{ 0.845_{ -0.035 }^{+0.044} }%
    {agegyr}{ 7.9 \pm 4.5  }%
    {vmag}{ 12.31 \pm 0.02 }
    {distance}{ 196 \pm 15  }%
  }[\PackageError{tree}{Undefined option to tree: #1}{}]%
}%

\newcommand{\ktwothirtytwocirc}[1]{%
  \IfEqCase{#1}{%
   {gamma_hires}{ -1.69 \pm 0.85  }%
   {gamma_pfs}{ -6.7 \pm 3.2  }%
   {jit_harps}{ 4.13 \pm 0.73  }%
   {k3}{ 2.3 \pm 1.1  }%
   {k2}{ 1.6 \pm 1.0  }%
   {k1}{ 5.63 \pm 0.91  }%
   {jit_hires}{ 3.77_{ -0.65 }^{+0.81} }%
   {jit_pfs}{ 6.5_{ -2.4 }^{+4.3} }%
   {gamma_harps}{ 1.07 \pm 0.84  }%
   {lnprobability}{ -237.5_{ -3.0 }^{+2.2} }%
   {e1}{  }%
   {e2}{  }%
   {e3}{  }%
   {P1}{ 8.99213 \pm 0.00016  }%
   {T01}{ 2076.91832 \pm 0.00055  }%
   {RpRstar1}{ 0.0556 \pm 0.0015  }%
   {Rp1}{ 5.13 \pm 0.28  }%
   {Mpsini1}{ 16.5 \pm 2.7  }%
   {rhop1}{ 0.67 \pm 0.16  }%
   {Teq1}{ 817 \pm 25  }%
   {Lstar1}{ 0.503_{ -0.052 }^{+0.067} }%
   {a1}{ 0.08036 \pm 0.00088  }%
   {Sinc1}{ 77.7_{ -8.3 }^{+10.8} }%
   {P2}{ 20.6602 \pm 0.0017  }%
   {T02}{ 2128.4067 \pm 0.0032  }%
   {RpRstar2}{ 0.0326 \pm 0.0023  }%
   {Rp2}{ 3.01 \pm 0.25  }%
   {Mpsini2}{ 6.2 \pm 3.9  }%
   {rhop2}{ 1.23 \pm 0.86  }%
   {Teq2}{ 619 \pm 19  }%
   {Lstar2}{ 0.503_{ -0.052 }^{+0.067} }%
   {a2}{ 0.1399 \pm 0.0015  }%
   {Sinc2}{ 25.6_{ -2.7 }^{+3.6} }%
   {P3}{ 31.7154 \pm 0.0022  }%
   {T03}{ 2070.7901 \pm 0.0026  }%
   {RpRstar3}{ 0.0371 \pm 0.0033  }%
   {Rp3}{ 3.43 \pm 0.35  }%
   {Mpsini3}{ 10.3 \pm 4.7  }%
   {rhop3}{ 1.38_{ -0.67 }^{+0.92} }%
   {Teq3}{ 537 \pm 16  }%
   {Lstar3}{ 0.503_{ -0.052 }^{+0.067} }%
   {a3}{ 0.1862 \pm 0.0020  }%
   {Sinc3}{ 14.5_{ -1.5 }^{+2.0} }%
   {kul2}{3.1} 
   {Mpsiniul2}{12.1} 
   {rhopul2}{2.7} 
  }[\PackageError{tree}{Undefined option to tree: #1}{}]%
}%

\newcommand{\ktwothirtytwoecc}[1]{%
  \IfEqCase{#1}{%
   {gamma_hires}{ -1.70 \pm 0.87  }%
   {sesinw1}{ 0.01 \pm 0.27  }%
   {gamma_pfs}{ -6.7 \pm 3.4  }%
   {jit_harps}{ 4.13 \pm 0.73  }%
   {k3}{ 2.4 \pm 1.1  }%
   {k2}{ 1.7 \pm 1.0  }%
   {k1}{ 5.60 \pm 0.93  }%
   {jit_hires}{ 3.88_{ -0.65 }^{+0.82} }%
   {jit_pfs}{ 6.6_{ -2.5 }^{+4.6} }%
   {secosw1}{ -0.08 \pm 0.20  }%
   {gamma_harps}{ 1.15 \pm 0.85  }%
   {lnprobability}{ -238.0_{ -3.3 }^{+2.3} }%
   {e1}{ 0.074_{ -0.055 }^{+0.093} }%
   {e2}{  }%
   {e3}{  }%
   {P1}{ 8.99213 \pm 0.00016  }%
   {T01}{ 2076.91832 \pm 0.00055  }%
   {RpRstar1}{ 0.0556 \pm 0.0015  }%
   {Rp1}{ 5.36 \pm 0.30  }%
   {Mpsini1}{ 16.5 \pm 2.8  }%
   {rhop1}{ 0.58_{ -0.13 }^{+0.16} }%
   {Teq1}{ 828 \pm 26  }%
   {Lstar1}{ 0.534 \pm 0.066  }%
   {a1}{ 0.08067_{ -0.00072 }^{+0.00089} }%
   {Sinc1}{ 82 \pm 10  }%
   {P2}{ 20.6602 \pm 0.0017  }%
   {T02}{ 2128.4067 \pm 0.0032  }%
   {RpRstar2}{ 0.0326 \pm 0.0023  }%
   {Rp2}{ 3.14 \pm 0.27  }%
   {Mpsini2}{ 6.6 \pm 4.0  }%
   {rhop2}{ 1.15 \pm 0.77  }%
   {Teq2}{ 628 \pm 20  }%
   {Lstar2}{ 0.534 \pm 0.066  }%
   {a2}{ 0.1405_{ -0.0012 }^{+0.0015} }%
   {Sinc2}{ 27.1 \pm 3.4  }%
   {P3}{ 31.7154 \pm 0.0022  }%
   {T03}{ 2070.7901 \pm 0.0026  }%
   {RpRstar3}{ 0.0371 \pm 0.0034  }%
   {Rp3}{ 3.57 \pm 0.37  }%
   {Mpsini3}{ 10.9 \pm 4.9  }%
   {rhop3}{ 1.29_{ -0.61 }^{+0.86} }%
   {Teq3}{ 544 \pm 17  }%
   {Lstar3}{ 0.534 \pm 0.066  }%
   {a3}{ 0.1869_{ -0.0017 }^{+0.0021} }%
   {Sinc3}{ 15.3 \pm 1.9  }%
   {kul2}{3.2}%
   {Mpsiniul2}{12.7}%
   {rhopul2}{2.5}%
   {eul1}{0.23}%
  }[\PackageError{tree}{Undefined option to tree: #1}{}]%
}%

\newcommand{\ktwothirtyninestar}[1]{%
  \IfEqCase{#1}{%
    {teff}{ 4912 \pm 60}
    {logg}{ 3.58 \pm 0.05}%
    {fe}{ 0.43 \pm 0.04}%
    {vsini}{ 0.1 }%
    {vmag}{ 10.83 \pm 0.08 }
    {mass}{ 1.192_{ -0.070 }^{+0.085} }
    {radius}{ 2.93 \pm 0.21  }%
    {agegyr}{ 6.7_{ -1.3 }^{+1.7} }%
    {plx}{ 3.50_{-0.20}^{+0.24}}%
    {distance}{ 267 \pm 24  }%
  }[\PackageError{tree}{Undefined option to tree: #1}{}]%
}%

\newcommand{\ktwothirtyninestarvaneylen}[1]{%
  \IfEqCase{#1}{%
    {teff}{ 4881 \pm 20}
    {logg}{ 3.44 \pm 0.07}%
    {fe}{ 0.32 \pm 0.04}%
    {mass}{ 1.35_{-0.03 }^{+0.04} }
    {radius}{ 3.90_{ -0.27 }^{ +0.30 } }%
    {agegyr}{ 3.09_{ -0.70 }^{ +0.92} }%
    {plx}{ 2.62 \pm 0.20 }%
  }[\PackageError{tree}{Undefined option to tree: #1}{}]%
}%

\newcommand{\ktwothirtyninecirc}[1]{%
  \IfEqCase{#1}{%
   {rrat1}{ 1.79 \pm 0.13}%
{jit_fies}{ 7.2 \pm 1.6  }%
{gamma_hires}{ 2.1 \pm 2.2  }%
{gamma_pfs}{ 1.3 \pm 3.6  }%
{jit_harps}{ 6.7 \pm 1.4  }%
{k2}{ 17.4 \pm 2.8  }%
{k1}{ 12.7 \pm 1.3  }%
{jit_hires}{ 7.42 \pm 0.86  }%
{jit_pfs}{ 5.8 \pm 1.4  }%
{gamma_fies}{ 2.7 \pm 3.1  }%
{tc2}{ 2456940 \pm 16  }%
{per2}{ 329 \pm 10  }%
{gamma_harps}{ -3.4 \pm 3.9  }%
{lnprobability}{ -268.1_{ -3.1 }^{+2.3} }%
{e1}{  }%
{e2}{  }%
{P1}{ 4.60497 \pm 0.00077  }%
{T01}{ 2152.4315 \pm 0.0058  }%
{RpRstar1}{ 0.0179 \pm 0.0014  }%
{Rp1}{ 5.71 \pm 0.63  }%
{Mpsini1}{ 37.3 \pm 4.3  }%
{rhop1}{ 1.10_{ -0.31 }^{+0.44} }%
{Teq1}{ 1670 \pm 54  }%
{Lstar1}{ 4.49 \pm 0.72  }%
{a1}{ 0.0574 \pm 0.0012  }%
{Sinc1}{ 1356 \pm 174  }%
{P2}{ 4.60497 \pm 0.00077  }%
{T02}{ 2152.4315 \pm 0.0058  }%
{RpRstar2}{ 0.0179 \pm 0.0015  }%
{Rp2}{ 5.71 \pm 0.63  }%
{Mpsini2}{ 51.1 \pm 8.7  }%
{rhop2}{ 1.50_{ -0.44 }^{+0.65} }%
{Teq2}{ 1670 \pm 54  }%
{Lstar2}{ 4.49 \pm 0.72  }%
{a2}{ 0.0574 \pm 0.0012  }%
{Sinc2}{ 1356 \pm 174  }%
  }[\PackageError{tree}{Undefined option to tree: #1}{}]%
}%

\newcommand{\ktwothirtynineecc}[1]{%
  \IfEqCase{#1}{%
{jit_fies}{ 6.7 \pm 1.6  }%
{gamma_hires}{ 1.9 \pm 2.2  }%
{sesinw1}{ -0.26_{ -0.17 }^{+0.25} }%
{gamma_pfs}{ -1.7 \pm 3.5  }%
{jit_harps}{ 6.3 \pm 1.4  }%
{k2}{ 18.4 \pm 2.9  }%
{k1}{ 13.8 \pm 1.4  }%
{jit_hires}{ 7.57 \pm 0.86  }%
{jit_pfs}{ 5.0 \pm 1.5  }%
{gamma_fies}{ 24574.5 \pm 3.0  }%
{secosw1}{ 0.261 \pm 0.086  }%
{tc2}{ 2456614 \pm 25  }%
{per2}{ 327.2 \pm 9.8  }%
{gamma_harps}{ 24486.6 \pm 3.9  }%
{lnprobability}{ -265.0_{ -3.4 }^{+2.6} }%
{e1}{ 0.152_{ -0.068 }^{+0.084} }%
{e2}{  }%
{P1}{ 4.60497 \pm 0.00077  }%
{T01}{ 2152.4315 \pm 0.0058  }%
{RpRstar1}{ 0.0179 \pm 0.0015  }%
{Rp1}{ 5.71 \pm 0.63  }%
{Mpsini1}{ 39.8 \pm 4.4  }%
{rhop1}{ 1.17_{ -0.32 }^{+0.47} }%
{Teq1}{ 1670 \pm 54  }%
{Lstar1}{ 4.48 \pm 0.72  }%
{a1}{ 0.0574 \pm 0.0012  }%
{Sinc1}{ 1356 \pm 175  }%
{P2}{ 4.60497 \pm 0.00077  }%
{T02}{ 2152.4315 \pm 0.0058  }%
{RpRstar2}{ 0.0179 \pm 0.0014  }%
{Rp2}{ 5.71 \pm 0.63  }%
{Mpsini2}{ 54.1 \pm 9.0  }%
{rhop2}{ 1.59_{ -0.47 }^{+0.68} }%
{Teq2}{ 1670 \pm 54  }%
{Lstar2}{ 4.48 \pm 0.72  }%
{a2}{ 0.0574 \pm 0.0012  }%
{Sinc2}{ 1356 \pm 175  }%
  }[\PackageError{tree}{Undefined option to tree: #1}{}]%
}%

\newcommand{\epictwentyoneseventeenstar}[1]{%
  \IfEqCase{#1}{%
    {name}{EPIC-211736671}%
    {teff}{ 5474 \pm 60}
    {logg}{ 3.99 \pm 0.05}%
    {fe}{ 0.33 \pm 0.04}%
    {vsini}{ < 2 }%
    {vmag}{ 12.33 \pm 0.01 }
    {mass}{ 1.121_{ -0.053 }^{+0.065} }
    {radius}{ 1.75 \pm 0.14  }%
    {agegyr}{ 7.8 \pm 1.5  }%
    {distance}{ 412 \pm 34  }%
  }[\PackageError{tree}{Undefined option to tree: #1}{}]%
}%

\newcommand{\epictwentyoneseventeencirc}[1]{%
  \IfEqCase{#1}{%
{gamma_hires}{ -4.0 \pm 1.6  }%
{k1}{ 19.4 \pm 2.3  }%
{jit_hires}{ 6.0_{ -1.1 }^{+1.5} }%
{lnprobability}{ -64.76_{ -1.80 }^{+0.92} }%
{e1}{  }%
{P1}{ 4.73401 \pm 0.00024  }%
{T01}{ 2312.0965 \pm 0.0019  }%
{RpRstar1}{ 0.0276 \pm 0.0018  }%
{Rp1}{ 5.28 \pm 0.54  }%
{Mpsini1}{ 55.1 \pm 6.8  }%
{rhop1}{ 2.05_{ -0.54 }^{+0.75} }%
{Teq1}{ 1446 \pm 47  }%
{Lstar1}{ 2.51 \pm 0.42  }%
{a1}{ 0.0573 \pm 0.0010  }%
{Sinc1}{ 762 \pm 100  }%
  }[\PackageError{tree}{Undefined option to tree: #1}{}]%
}%

\newcommand{\epictwentyoneseventeenecc}[1]{%
  \IfEqCase{#1}{%
{secosw1}{ -0.324_{ -0.061 }^{+0.076} }%
{dvdt}{ -11.0 \pm 2.3  }%
{gamma_hires}{ -2.63 \pm 0.91  }%
{k1}{ 21.3 \pm 1.4  }%
{jit_hires}{ 2.86_{ -0.78 }^{+1.01} }%
{sesinw1}{ 0.26_{ -0.14 }^{+0.11} }%
{lnprobability}{ -52.6_{ -2.8 }^{+1.7} }%
{e1}{ 0.180 \pm 0.042  }%
{w1}{ 2.44 \pm 0.33  }%
{ecosw1}{ -0.135 \pm 0.034  }%
{esinw1}{ 0.107 \pm 0.064  }%
{P1}{ 4.73401 \pm 0.00024  }%
{T01}{ 2312.0965 \pm 0.0019  }%
{RpRstar1}{ 0.0276 \pm 0.0018  }%
{Rp1}{ 5.28 \pm 0.54  }%
{Mpsini1}{ 59.4 \pm 4.4  }%
{rhop1}{ 2.22_{ -0.55 }^{+0.77} }%
{Teq1}{ 1446 \pm 48  }%
{Lstar1}{ 2.50 \pm 0.42  }%
{a1}{ 0.0573 \pm 0.0010  }%
{Sinc1}{ 762 \pm 100  }%
  }[\PackageError{tree}{Undefined option to tree: #1}{}]%
}%

\newcommand{\val}[2]{%
  \IfEqCase{#1}{%
    {K2-27-star}{\ktwotwentysevenstar{#2}}%
    {K2-27-circ}{\ktwotwentysevencirc{#2}}%
    {K2-27-ecc}{\ktwotwentysevenecc{#2}}%
    {K2-32-star}{\ktwothirtytwostar{#2}}%
    {K2-32-circ}{\ktwothirtytwocirc{#2}}%
    {K2-32-ecc}{\ktwothirtytwoecc{#2}}%
    {K2-39-star}{\ktwothirtyninestar{#2}}%
    {K2-39-star-VE}{\ktwothirtyninestarvaneylen{#2}}%
    {K2-39-circ}{\ktwothirtyninecirc{#2}}%
    {K2-39-ecc}{\ktwothirtynineecc{#2}}%
    {epic2117-star}{\epictwentyoneseventeenstar{#2}}%
    {epic2117-circ}{\epictwentyoneseventeencirc{#2}}%
    {epic2117-ecc}{\epictwentyoneseventeenecc{#2}}%
  }[\PackageError{tree}{Undefined option to tree: #1}{}]%
}%

\begin{document}


\title{Four Sub-Saturns with Dissimilar Densities: Windows into Planetary Cores and Envelopes}



\author{Erik A. Petigura\altaffilmark{1,2,13}}
\author{Evan Sinukoff\altaffilmark{3,14}}
\author{Eric Lopez\altaffilmark{4}}
\author{Ian J. M. Crossfield\altaffilmark{5,15}}
\author{Andrew W. Howard\altaffilmark{1}}
\author{John M. Brewer\altaffilmark{6}}
\author{Benjamin J. Fulton\altaffilmark{1,3}}
\author{Howard T. Isaacson\altaffilmark{7}}
\author{David R. Ciardi\altaffilmark{8}}
\author{Steve B. Howell\altaffilmark{9}}
\author{Mark E. Everett\altaffilmark{10}}
\author{Elliott P. Horch\altaffilmark{11}}
\author{Lea Hirsch\altaffilmark{7}}
\author{Lauren M. Weiss\altaffilmark{12,16}}
\and
\author{Joshua E. Schlieder\altaffilmark{8}}


\altaffiltext{1}{California Institute of Techonology}
\altaffiltext{2}{petigura@caltech.edu}
\altaffiltext{3}{Institute for Astronomy, University of Hawai`i at M\={a}noa} 
\altaffiltext{4}{Institute for Astronomy, University of Edinburgh}
\altaffiltext{5}{University of California, Santa Cruz}
\altaffiltext{6}{Yale University}
\altaffiltext{7}{University of California, Berkeley}
\altaffiltext{8}{IPAC-NExScI, California Institute of Technology}
\altaffiltext{9}{NASA Ames Research Center}
\altaffiltext{10}{National Optical Astronomy Observatory}
\altaffiltext{11}{Department of Physics, Southern Connecticut State University}
\altaffiltext{12}{Institut de Recherche sur les Exoplan\`etes, Universit\'e de Montr\'eal}
\altaffiltext{13}{Hubble Fellow}
\altaffiltext{14}{NSERC Fellow}
\altaffiltext{15}{NASA Sagan Fellow}
\altaffiltext{16}{Trottier Fellow}

\begin{abstract}
We present results from a Keck/HIRES radial velocity campaign to study four sub-Saturn-sized planets, K2-27b, K2-32b, K2-39b, and K2-108b, with the goal of understanding their masses, orbits, and heavy element enrichment. The  planets have similar sizes (\Rp =  4.5--5.5~\Re), but have dissimilar masses (\Mp~=~16--60~\Me), implying a diversity in their core and envelope masses. K2-32b is the least massive ($\Mp = \val{K2-32-circ}{Mpsini1}~\Me$) and orbits in close proximity to two sub-Neptunes near a 3:2:1 period commensurability. K2-27b and K2-39b are significantly more massive at $\Mp = \val{K2-27-ecc}{Mpsini1}~\Me$ and $\Mp = \val{K2-39-ecc}{Mpsini1}~\Me$, respectively, and show no signs of additional planets. K2-108b is the most massive at $\Mp = \val{epic2117-ecc}{Mpsini1}~\Me$, implying a large reservoir of heavy elements of about $\approx50$~\Me. Sub-Saturns as a population have a large diversity in planet mass at a given size. They exhibit remarkably little correlation between mass and size; sub-Saturns range from $\approx $6--60~\Me, regardless of size. We find a strong correlation between planet mass and host star metallicity, suggesting that metal-rich disks form more massive planet cores. The most massive sub-Saturns tend to lack detected companions and have moderately eccentric orbits, perhaps as a result of a previous epoch of dynamical instability. Finally, we observe only a weak correlation between the planet envelope fraction and present-day equilibrium temperature, suggesting that photo-evaporation does not play a dominant role in determining the amount of gas sub-Saturns accrete from their protoplanetary disks.
\end{abstract}

\section{Introduction}
The Solar System contains four terrestrial planets, two ice giants, and two gas giants on nearly circular and co-planar orbits. Notably, the Solar System lacks several broad classes of planets: it contains no planets having sizes between Earth and Neptune (1.0--3.9~\Re) or between Uranus and Saturn (4.0--9.4~\Re), and no planets that orbit closer than Mercury (0.39~\AU). A longstanding question is whether the Solar System is representative of planetary systems around other stars, or if it is one particular realization of a set of physical processes that produce a diversity of outcomes.

The study of extrasolar planets offers a path to address this question. Studies of planet occurrence from the prime \Kepler mission \citep{Borucki10a} revealed that our Solar System is atypical in a few key ways: the majority of stars have at least one planet interior to Mercury's orbit and the most common size of planet is in the range 1--3~\Re, sizes not represented in our Solar System \citep{Howard12,Fressin13,Petigura13b}. The occurrence of planets rises rapidly below \Rp~=~3.0~\Re, indicating an important size scale in the formation of planet cores and envelopes. Constraining the bulk composition of these sub-Neptunes has been the focus of intensive radial velocity (RV) campaigns that revealed that most planets larger than $\approx 1.6~\Re$ have significant gaseous envelopes \citep{Marcy14,Weiss14,Rogers15}.

In this paper, we focus on another size class of planets absent in the Solar System, sub-Saturns, which we define as planets having sizes between 4.0--8.0~\Re. Sub-Saturns offer a superb laboratory to study planet formation history and compositions. Their large sizes require significant envelopes of H/He. For sub-Saturns, H/He comprise such a large component of the planet volume that the planets can be modeled as a high-density core with a thick H/He envelope. Measurements of the core mass fraction are simplified because details in the composition of the core have little effect on the measured planet size \citep{Lopez14,Petigura16}.

While close-in (< 1 AU) gaseous planets are thought to form via core accretion \citep{Pollack96,Bodenheimer00,Hubickyj05,Mordasini08} there are major uncertainties regarding how planets acquire (and lose) mass and the extent to which their orbits evolve with time. For example, Jupiter is thought to have taken $\sim$3 Myr to accrete enough gas before triggering runaway accretion \citep{Hubickyj05} while gas disks around other stars are observed to dissipate after 1--10~Myr \citep{Mamajek09}. That these two processes have similar timescales may explain why the occurrence of Jovians within 20~AU is $\approx20\%$ rather than $\approx 100\%$ \citep{Cumming08}.

While sub-Saturns are similar to Jovians given their H/He envelopes, they often have much lower masses. For example, Kepler-79d is 7.2~\Re and only 6.0~\Me \citep{Jontof-Hutter14}. For sub-Saturns, the runaway accretion of gas invoked to explain Jupiter's massive envelope seems to have not occurred. Low-density sub-Saturns have inspired alternative gas accretion scenarios, such as accretion in a gas-depleted disk (e.g. \citealt{Lee15b}).

Here, we present RV measurements of four Sub-Saturns, K2-27b, K2-32b, K2-39b, and K2-108b, taken as part of a program to expand the sample of sub-Saturns with well-measured masses and radii. These planets were observed by the {\em Kepler Space Telescope} \citep{Borucki10a} operating during its \ktwo mission, where the telescope observes a new field in the ecliptic plane every $\approx3$~months \citep{howell:2014}. Sections \ref{sec:rv}-\ref{sec:individual} present the radial velocity measurements, stellar characterization, and modeling needed to extract planet mass, radius, and orbital eccentricity. For these planets, we achieve mass measurements of 16\% or better and density measurements to 33\% or better. 

In Section~\ref{sec:discussion}, we place the four planets in the context of other sub-Saturns. We find a large diversity in planet masses in the sub-Saturn size range with little correlation with planet size. Sub-Saturns range from $\approx$6--60~\Me, regardless of their size. We find a strong correlation between stellar metallicity and planet mass, suggesting metal-rich disks likely form more massive planet cores. We also observe that planet mass seems to be inversely correlated with the presence of additional planets, which could be the result of  a period of large-scale dynamical instabilities resulting in mergers or scattering on to high inclination orbits. We apply the interior structure models of \citep{Lopez14} to determine the fraction of planet mass in H/He and heavy element. For sub-Saturns, we see only a weak dependence of the envelope fraction on equilibrium temperature, indicating that photo-evaporation does not likely play a major role in sculpting the final sizes of sub-Saturns. We offer some concluding thoughts in Section~\ref{sec:conclusions}.

\section{Radial Velocity Observations and Analysis}
\label{sec:rv}
Here, we describe our overall RV observational campaign and our analysis methodology. Details on individual systems are given in  Section~\ref{sec:individual}. We observed K2-27, K2-32, K2-39, and K2-108 using the High Resolution Echelle Spectrometer (HIRES; \citealt{Vogt94}) on the 10~m Keck Telescope I. We collected spectra through an iodine cell mounted directly in front of the spectrometer slit. The iodine cell imprints a dense forest of absorption lines which serve as a wavelength reference. We used an exposure meter to achieve a consistent signal to noise level for each program star, which ranged from 100 to 130 per reduced pixel on blaze near 550~nm. We also obtained a ``template'' spectrum without iodine. 

RVs were determined using standard procedures of the California Planet Search \citep[CPS;][]{howard:2010b} including forward modeling of the stellar and iodine spectra convolved with the instrumental response \citep{Marcy92,Valenti95}. The RVs are tabulated in Table~\ref{tab:rv}. We also list the measurement uncertainty of each RV point, which ranges from 1.5 to 2.0~\ms and is derived from the uncertainty on the mean RV of the $\sim$700 spectral chunks used in the RV pipeline. 

We analyzed the RV time-series using the publicly-available RV-fitting package \texttt{radvel} (Fulton \& Petigura, in prep.)%
\footnote{\url{https://github.com/California-Planet-Search/radvel}}
When modelling the RVs, we adopt the likelihood, $\mathcal{L}$, of \citet{Howard14}:
\[
    \ln{\mathcal{L}} = 
        - \frac{1}{2}\sum_{i} 
        \left[
            \frac{\left(v_i-v_{m,i}\right)^2}{\sigma_i^2+ \sigjit{}^2}
            + \ln{2\pi\left(\sigma_i^2 + \sigjit{}^2\right)}
        \right],
\]
where $v_i$ is the \textit{i}'th RV measured at time $t_i$, $\sigma_i$ is the corresponding uncertainty, $v_m$ is the Keplerian model velocity at $t_i$, and \sigjit{} or ``jitter'' accounts for additional RV variability due to stellar and instrumental noise and is included in our models as a free parameter. 

When modeling the RVs, we first consider circular Keplerians with no additional acceleration term, \dvdt. Here, \sigjit{} and an average RV offset, \gam{}, are allowed to float as free parameters. We then allow for more complicated models if they are motivated by the data. We consider models where  \dvdt is allowed to float and models where eccentricity, $e$, and longitude of periastron, \lonperi, are allowed to vary.%
\footnote{
During fitting and MCMC modeling, we parametrize $e$ and \lonperi by \sqrtecosw and \sqrtesinw, as recommended by \cite{Eastman13}, to guard against the Lucy-Sweeney bias toward non-zero eccentricities.
}
More complex models will naturally lead to higher likelihoods at the expense of additional free parameters. To assess whether a more complex model is justified, we use the Bayesian Information Criterion (BIC; \citealt{Schwartz78}). Models with smaller BIC, i.e. negative \dbic are preferend.  

When available, we incorporated RV measurements from the literature to augment our HIRES timeseries. RVs of K2-27, K2-32, and K2-39 have been published in \cite{VanEylen16a}, \cite{Dai16}, and \cite{VanEylen16b}, respectively.  We fit for the offset and jitter terms independently for different datasets. To derive uncertainties on the RV parameters we perform a standard MCMC exploration of the likelihood surface using the {\tt emcee} Python package \citep{Goodman10,Foreman-Mackey13}. In Tables~\ref{tab:k2-27}--\ref{tab:epic2117}, we list orbital and planetary properties assuming both circular and eccentric orbits. The preferred model is indicated.

\begin{deluxetable}{llRRR}
\tablecaption{Radial Velocities\label{tab:rv}}
\tablecolumns{5}
\tablewidth{-0pt}
\tabletypesize{\footnotesize}
\tablehead{
        \colhead{Star} & 
        \colhead{Inst.} & 
        \colhead{Time} & 
        \colhead{RV} & 
        \colhead{$\sigma$(RV)}\\
        \colhead{} & 
        \colhead{} & 
        \colhead{\bjdtdb} & 
        \colhead{\ms} & 
        \colhead{\ms}
        }
\startdata
K2-27 & HIRES & 2457059.023437 & -3.30 & 3.58 \\
K2-27 & HARPS & 2457187.504940 & -37782.19 & 2.56 \\
K2-27 & HARPS-N & 2457064.713740 & -37785.55 & 6.56 \\
K2-27 & FIES & 2457045.607900 & -38039.00 & 11.10 \\
K2-32 & HIRES & 2457179.918605 & 2.24 & 1.81 \\
K2-32 & HARPS & 2457185.606900 & 10.69 & 2.65 \\
K2-32 & PFS & 2457198.674600 & -13.95 & 2.31 \\
K2-39 & HIRES & 2457245.118029 & -2.77 & 1.84 \\
K2-39 & HARPS & 2457255.714330 & 24507.93 & 2.66 \\
K2-39 & PFS & 2457257.799090 & 0.73 & 1.64 \\
K2-39 & FIES & 2457235.669620 & 24557.22 & 6.99 \\

\enddata
\tablecomments{The radial velocity (RV) measurements used in this work. We list the HIRES RVs along with other RVs from the literature, where available. Table \ref{tab:rv} is published in its entirety in machine-readable format. A portion is shown here for guidance regarding its form and content.}
\end{deluxetable}

\section{Stellar Properties}
\label{sec:stellar}
We measured stellar effective temperature, \teff, surface gravity, \logg, and metallicity, \fe from our iodine-free ``template'' spectra. We followed the methodology of \cite{Brewer16}, which used an updated version of the \SME code and has been shown to recover surface gravities consistent with those from asteroseismology to within 0.05 dex \citep{Brewer15}. We constrained stellar mass, \Mstar, radius, \Rstar, and age from \teff, \logg, and \fe using the {\tt isochrones} Python package \citep{Morton15} which interpolates among the Dartmouth stellar isochrones \citep{Dotter08}. The stellar properties are listed alongside other system properties in Tables~\ref{tab:k2-27}-\ref{tab:epic2117}, respectively. 

For K2-39, we note some tension between the stellar parameters presented by \cite{VanEylen16b} and those presented here. Importantly, the \logg derived by \cite{VanEylen16b} is lower than the \SME value. The \cite{VanEylen16b} analysis resulted in a larger inferred stellar radius measurement that, at \Rstar~=~$\val{K2-39-star-VE}{radius}~$\Rsun, is 34\% larger than the \Rstar~=~$\val{K2-39-star}{radius}~$\Rsun derived by \SME and {\tt isochrones}. We present a side-by-side comparison of spectroscopic parameters in Table~\ref{tab:k2-39-stellar}. 

Can the recently released {\em Gaia} parallaxes be used to resolve these different estimates of stellar radius? K2-39 is listed in the Tycho catalog as TYC-5811-835-1. \Gaia recently released parallaxes for most stars in the 2 million-star Tycho catalog. K2-39 has a parallax of $3.35\pm0.86$~mas.  Given the apparent $K$-band magnitude of $8.516 \pm 0.024$ from 2MASS \citep{Cutri03}, we calculated the parallax implied by both sets of spectroscopic parameters. We used $K$-band since it is less sensitive to the unknown amount of extinction between Earth and K2-39. When we used the \SME parameters we found parallax of $\val{K2-39-star}{plx}$~mas, in close agreement with the \Gaia parallax. The larger radius of \cite{VanEylen16b} necessitates a more distant star to produce the observed $K$~mag, and thus yields a smaller expected parallax of $\val{K2-39-star-VE}{plx}$~mas. While this parallax is smaller than the most likely \Gaia value, it is still consistent at the 1$\sigma$ level. Given that \Gaia Data Release 1 provided a 4$\sigma$ measurement of K2-39's parallax, it is insufficient to distinguish between the two sets of parameters. Future \Gaia releases will provide important constraints on the physical properties of K2-39. 

\cite{VanEylen16b} also noted that the non-detection of asteroseismic modes places a lower limit of $\logg \geq 3.50$~dex. Taken together, the non-dection of asteroseismic modes and the {\em Gaia} parallax both suggest that K2-39 is smaller than reported in \cite{VanEylen16b}. Given that \SME has been extensively tested against asteroseismology we adopt the \SME parameters hereafter. 

\begin{deluxetable}{lrr}
\tablecaption{K2-39 Stellar Parameters\label{tab:k2-39-stellar}}
\tablecolumns{3}
\tablehead{\colhead{} & \colhead{This work} & \colhead{V16}}
\tablewidth{0pt}
\startdata
\teff (K)      & $\val{K2-39-star}{teff}$   & $\val{K2-39-star-VE}{teff}$   \\
\logg (dex)    & $\val{K2-39-star}{logg}$   & $\val{K2-39-star-VE}{logg}$   \\
\fe (dex)      & $\val{K2-39-star}{fe}$     & $\val{K2-39-star-VE}{fe}$     \\
\Mstar (\Msun) & $\val{K2-39-star}{mass}$   & $\val{K2-39-star-VE}{mass}$   \\
\Rstar (\Rsun) & $\val{K2-39-star}{radius}$ & $\val{K2-39-star-VE}{radius}$ \\
Parallax (mas)\tablenotemark{a} & $\val{K2-39-star}{plx}$    & $\val{K2-39-star-VE}{plx}$    \\
\enddata
\tablecomments{Comparison of the stellar properties from this work (based on \SME, \citealt{Brewer15}), and from \cite{VanEylen16b}. \Mstar and \Rstar in the \cite{VanEylen16b} column were derived using the {\tt isochrones} package which interpolates between Dartmouth isochrones, as opposed to \cite{VanEylen16b} who used Yonsei-Yale isochrones. However, \cite{VanEylen16b} gives \Rstar = $3.88^{+0.48}_{-0.42}$~\Rsun, which is consistent with the Dartmouth model. We adopt the \SME value given that the code has been extensively tested against asteroseismology.}
\tablenotetext{a}{Implied parallax from based on spectroscopic properties, isochrone modeling, and apparent $K$-band magnitude.}
\end{deluxetable}

\section{Individual Targets}
\label{sec:individual}
\subsection{K2-27}
\label{ssec:k2-27}
K2-27 hosts a single transiting sub-Saturn, K2-27b, with $P = 6.77$~d that was first confirmed in \cite{VanEylen16a} using RVs. We obtained 15 spectra with HIRES of K2-27 between 2015-02-05 and 2016-07-17. \cite{VanEylen16a} observed this star with HARPS, HARPS-N, and FIES. We included 6 and 19 measurements from HARPS and HARPS-N, respectively in our RV analysis. We did not include the 6 FIES measurements because their uncertainties are much larger ($\approx 30$~\ms), compared to uncertainties from the HIRES, HARPS, and HARPS-N spectra ($\approx 4-5$~\ms) and hence add little additional information to constrain the RV fits while increasing model complexity.

\begin{figure*}
\centering
\includegraphics[width=0.8\textwidth]{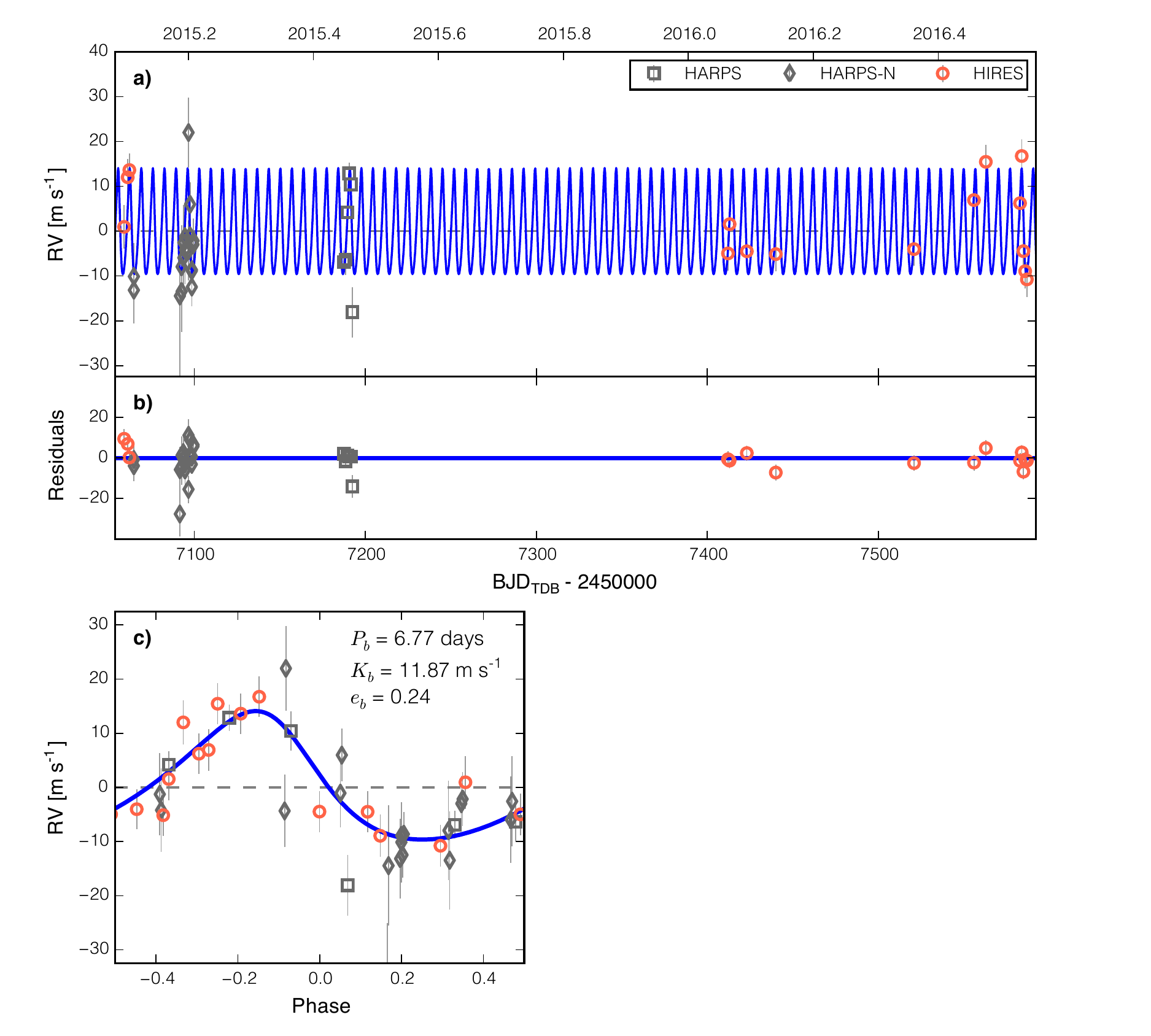}
\caption{Single Keplerian model of K2-27 radial velocities (RVs), allowing for eccentricity (see Section~\ref{ssec:k2-27}). {\bf a)} Time series of RVs from HIRES along with HARPS and HARPS-N published in \cite{VanEylen16a}. During the fitting, we allowed for an arbitrary offset between the three instruments to float as a free parameter. The blue line shows the most probable Keplerian model. {\bf b)} Residuals to the most probable Keplerian model. {\bf c)} Phase-folded RVs and the most probable Keplerian.\label{fig:k2-27}}
\end{figure*}

We first modeled the combined RVs assuming circular orbits and no additional acceleration term, \dvdt. We then allowed \dvdt to vary, but found that these models produced a negligible improvement in the BIC ($\dbic=-1$). We then allowed for eccentricity to float and found a non-zero eccentricity of $e = \val{K2-27-ecc}{e1}$. Compared to the circular models, the eccentric fits were strongly favored ($\dbic=-14$). We verified that the eccentricity is detected in both HIRES and HARPS+HARPS-N datasets by fitting these subsets independently. These datasets yield consistent and non-zero eccentricities. We adopted the parameters from the eccentric model as the preferred parameters. The most probable eccentric model is shown in Figure~\ref{fig:k2-27}. The K2-27 system parameters are summarized in Table~\ref{tab:k2-27}. We measured a Doppler semi-amplitude of $\K{} = \val{K2-27-ecc}{k1}$~\ms, which is consistent with $\K{} = 10.8 \pm 2.7$~\ms reported by \cite{VanEylen16a}, but with smaller uncertainties due to our additional measurements. We measured a mass of $\Mp = \val{K2-27-ecc}{Mpsini1}$~\Me and a density of $\val{K2-27-ecc}{rhop1}$~\gcc. As we discuss in Section~\ref{sec:discussion}, K2-27b has a relatively high mass for its size, implying a large core-mass.

{\renewcommand{\arraystretch}{0.9}
\begin{table}
\centering
\caption{System parameters of K2-27}
\begin{tabular}{lrr}
\hline
\hline
   & Value   & Ref.  \\
    \hline
    \multicolumn{3}{l}{{\bf Stellar parameters}} \\
    Identifier          & EPIC-201546283                & \\
    \teff (K)           & $\val{K2-27-star}{teff}$      & A \\
    \logg (dex)         & $\val{K2-27-star}{logg}$      & A \\
    \fe (dex)           & $\val{K2-27-star}{fe}$        & A \\
    \vsini (\kms)       & $\val{K2-27-star}{vsini}$     & A \\
    \Mstar (\Msun)      & $\val{K2-27-star}{mass}$      & A \\
    \Rstar (\Rsun)      & $\val{K2-27-star}{radius}$    & A \\
    age (Gyr)           & $\val{K2-27-star}{agegyr}$    & A \\
    Apparent $V$ (mag)  & $\val{K2-27-star}{vmag}$      & A \\
        \\[-2ex]
                        & {\bf planet b}                &  \\ 
    \multicolumn{3}{l}{{\bf Transit model}} \\
    $P$ (days)          & $\val{K2-27-ecc}{P1}$    & B\\
    $T_0$ (BJD-2454833) & $\val{K2-27-ecc}{T01}$   & B\\
    \Rp (\Re)           & $\val{K2-27-ecc}{Rp1}$   & A\\
        $a$ (AU)        & $\val{K2-27-ecc}{a1}$    & A\\
        \Sinc (\Se)     & $\val{K2-27-ecc}{Sinc1}$ & A\\
        \Teq (K)        & $\val{K2-27-ecc}{Teq1}$  & A\\
        \\[-2ex]
    \multicolumn{3}{l}{{\bf Circular RV model}} \\
    $K$ (\ms)              & $\val{K2-27-circ}{k1}$            & A \\
    \gam{HIRES} (\ms)      & $\val{K2-27-circ}{gamma_hires}$   & A \\
    \gam{HARPS} (\ms)      & $\val{K2-27-circ}{gamma_harps}$   & A \\
    \gam{HARPS-N} (\ms)    & $\val{K2-27-circ}{gamma_harps-n}$ & A \\
    \dvdt (\msyr)          & 0 (fixed)                         & A \\
    \sigjit{HIRES} (\ms)   & $\val{K2-27-circ}{jit_hires}$     & A \\
    \sigjit{HARPS} (\ms)   & $\val{K2-27-circ}{jit_harps}$     & A \\
    \sigjit{HARPS-N} (\ms) & $\val{K2-27-circ}{jit_harps-n}$   & A \\
    \Mp (\Me)              & $\val{K2-27-circ}{Mpsini1}$       & A \\
    $\rho$ (\gcc)          & $\val{K2-27-circ}{rhop1}$         & A \\
        \\[-2ex]
    \multicolumn{3}{l}{{\bf Eccentric RV model (adopted)}} \\
    $K$ (\ms)              & $\val{K2-27-ecc}{k1}$             & A \\
    $e$                    & $\val{K2-27-ecc}{e1}$             & A \\
    \gam{HIRES} (\ms)      & $\val{K2-27-ecc}{gamma_hires}$    & A \\
    \gam{HARPS} (\ms)      & $\val{K2-27-ecc}{gamma_harps}$    & A \\
    \gam{HARPS-N} (\ms)    & $\val{K2-27-ecc}{gamma_harps-n}$  & A \\
    \dvdt (\msyr)          & 0 (fixed)                         & A \\
    \sigjit{HIRES} (\ms)   & $\val{K2-27-ecc}{jit_hires}$      & A \\
    \sigjit{HARPS} (\ms)   & $\val{K2-27-ecc}{jit_harps}$      & A \\
    \sigjit{HARPS-N} (\ms) & $\val{K2-27-ecc}{jit_harps-n}$    & A \\
    \Mp (\Me)              & $\val{K2-27-ecc}{Mpsini1}$        & A \\
    $\rho$ (\gcc)          & $\val{K2-27-ecc}{rhop1}$          & A \\
\hline
\end{tabular}
\tablecomments{A: This work; B: \cite{Crossfield16}}
\label{tab:k2-27}
\end{table}
}

\subsection{K2-32}
\label{ssec:k2-32}
K2-32 hosts three planets, K2-32b, K2-32c, and K2-32d, having orbital periods of $P$ = 8.99~d, 20.66~d, and 31.7~d, respectively, which are near the 3:2:1 period commensurability. The planets were first confirmed in \cite{Sinukoff16} using multiplicity arguments \citep{Lissauer12}. We obtained 31 spectra of K2-32 with HIRES between 2015-06-06 and 2016-08-20. \cite{Dai16} obtained 43 spectra with HARPS and 6 with PFS, which we included in our RV analysis. We first modeled the combined HIRES, HARPS, and PFS RVs assuming circular orbits and no additional acceleration term, \dvdt. When allowing \dvdt to float, we found that $\dvdt = 2.3 \pm 1.8 $~\msyr, consistent with zero at the 2$\sigma$ level. This differs from $\dvdt = 34.0^{+9.9}_{-9.7}$~\msyr reported by \cite{Dai16}. All but 5 of the RVs from \cite{Dai16} were collected within a 25-day window, and thus do not provide much leverage on \dvdt. The positive \dvdt measured by \cite{Dai16} is driven by two RV measurements that fall above our best fit curve. With our combined dataset spanning two observing seasons, we place tighter limits on \dvdt and on the presence of long period companions having $P \gtrsim 2$~years. Nonetheless, the BIC preferred fixing \dvdt at 0~\msyr ($\dbic = 1$).

\begin{figure*}
\plotone{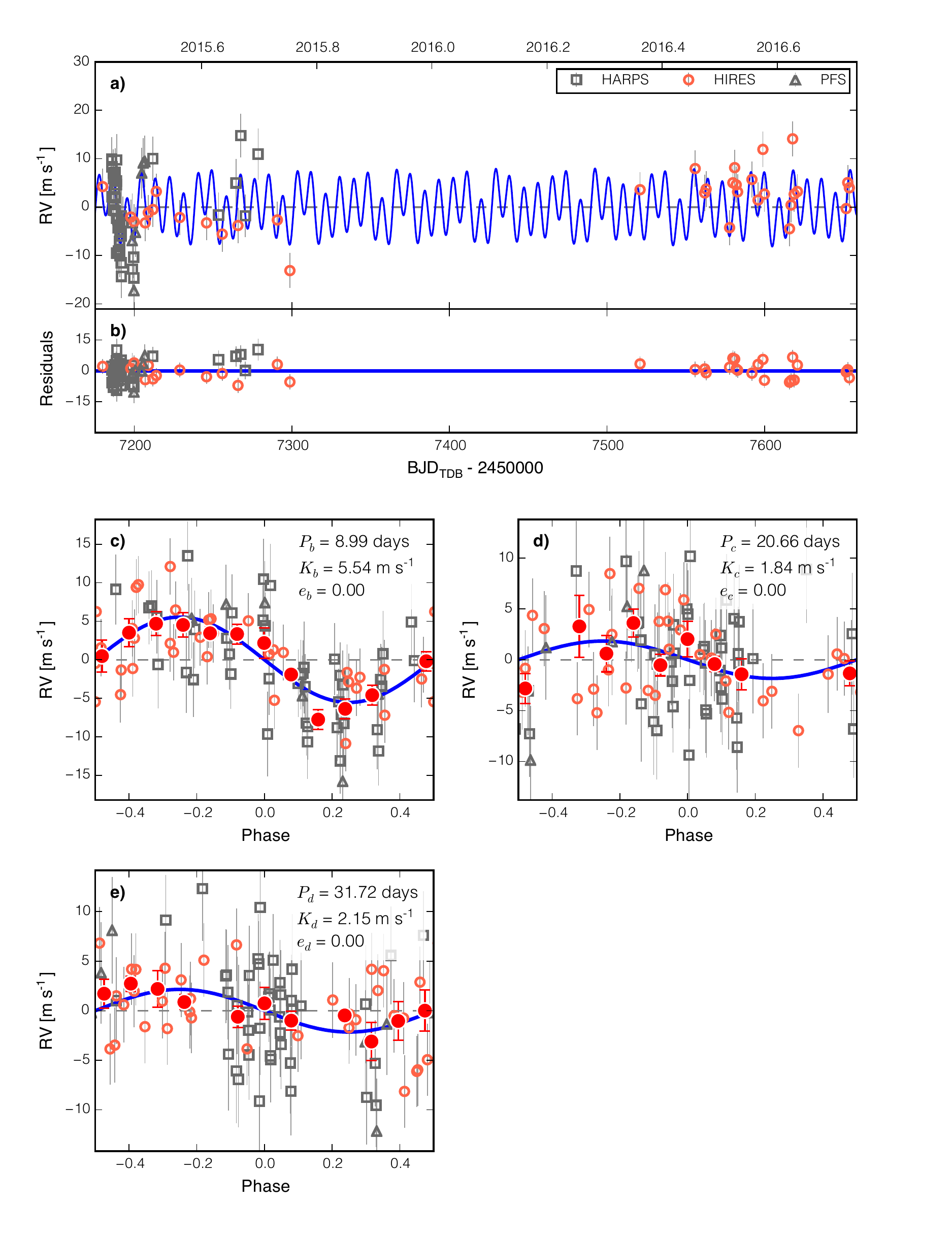}
\caption{Three-planet Keplerian fit to the K2-32 radial velocities (RVs), assuming circular orbits (see Section~\ref{ssec:k2-32}). {\bf a)} Time series of RVs from HIRES along with HARPS and PFS published in \cite{Dai16}. During the fitting, we allowed for an arbitrary offset between the three instruments to float as a free parameter. The blue line shows the most probable Keplerian model. {\bf b)} Residuals to the most probable Keplerian model. Panels {\bf c)} through {\bf e)} show the phase-folded RVs and the most probable Keplerian with the contributions from the other planets removed. The large red circles show the phase-binned RVs \label{fig:k2-32}.}
\end{figure*}

Next, we considered eccentric models. In the circular models, $\K{b} = \val{K2-32-circ}{k2}$~\ms and $\K{c} = \val{K2-32-circ}{k3}$~\ms, respectively. Given that the reflex motion due to K2-32c and K2-32d were only detected at the 2$\sigma$ level, we cannot place meaningful constraints on their eccentricities. Therefore, we fixed their eccentricities at zero in our fits and allowed the eccentricity of K2-32b to float. We found that the eccentricity of K2-32b was consistent with zero, and placed an upper limit of $\e{b} < \val{K2-32-ecc}{eul1}$ (95\% conf.). We therefore adopted the circular model for the K2-32 system parameters. The physical properties of the K2-32 system along with our circular and eccentric models are listed Table~\ref{tab:k2-32}. The best-fit circular model is shown in Figure~\ref{fig:k2-32}.

For K2-32b, we measured $\Mp = \val{K2-32-circ}{Mpsini1}$~\Me, which is lower than, but within the $1\sigma$ confidence interval of \Mp = $21.1 \pm 5.9$, measured by \cite{Dai16}. We measure a density of $\rho$ = $\val{K2-32-circ}{rhop1}$~\gcc, which is low compared to other sub-Saturns of similar size with RV mass measurements. However, when compared to the ensemble of mass measurements from TTVs and RVs, K2-32b is of intermediate density (see Section~\ref{sec:discussion}). 

Figure~\ref{fig:k2-32} shows that the \cite{Dai16} measurements were not well-sampled during the quadrature times of K2-32c and K2-32d, leading to poor constraints on their Doppler semi-amplitudes with significant covariance between \K{c} and \K{d} due the 3:2 period ratios of the planets. Our combined dataset has more uniform sampling in phase with minimal covariance between \K{c} and \K{d}.

We achieve marginal detections of K2-32c and K2-32d, which have masses of $\val{K2-32-circ}{Mpsini2}$~\Me and $\val{K2-32-circ}{Mpsini3}$~\Me, respectively. Because K2-32c is not quite a 2$\sigma$ detection, we conservatively report an upper limit of \Mp < $\val{K2-32-circ}{Mpsiniul2}$~\Me (95\% conf.), to be conservative. Continued Doppler monitoring of K2-32 is necessary to place tighter constraints on the masses and orbits of K2-32c and K2-32d. While such observations are challenging with current instruments given the faint host star ($V = 12.3$~mag), the dynamical richness of this system makes these observations worthwhile.

{\renewcommand{\arraystretch}{0.9}
\begin{table*}[htbp!]
\centering
\caption{K2-32 System Parameters}
\begin{tabular}{lrrrr}
\hline
\hline
 & Value  & & & Ref.  \\
\hline
    \multicolumn{5}{l}{{\bf Stellar parameters}} \\
    Identifier         & EPIC-205071984             &  & &   \\
    \teff (K)          & $\val{K2-32-star}{teff}$   &  & & A \\
    \logg (dex)        & $\val{K2-32-star}{logg}$   &  & & A \\
    \fe (dex)          & $\val{K2-32-star}{fe}$     &  & & A \\
    \vsini (\kms)      & $\val{K2-32-star}{vsini}$  &  & & A \\
    \Mstar (\Msun)     & $\val{K2-32-star}{mass}$   &  & & A \\
    \Rstar (\Rsun)     & $\val{K2-32-star}{radius}$ &  & & A \\
    age (Gyr)          & $\val{K2-32-star}{agegyr}$ &  & & A \\
    Apparent $V$ (mag) & $\val{K2-32-star}{vmag}$   &  & & A \\
        \\[-2ex]
    {}                  & {\bf planet b} & {\bf planet c} & {\bf planet d} &  \\ 
    \multicolumn{5}{l}{{\bf Transit model}} \\
    $P$ (days)          & $\val{K2-32-circ}{P1}$    & $\val{K2-32-circ}{P2}$     & $\val{K2-32-circ}{P3}$      & B\\
    $T_0$ (BJD-2454833) & $\val{K2-32-circ}{T01}$   & $\val{K2-32-circ}{T02}$    & $\val{K2-32-circ}{T03}$     & B\\
    \Rp (\Re)           & $\val{K2-32-circ}{Rp1}$   & $\val{K2-32-circ}{Rp2}$    & $\val{K2-32-circ}{Rp3}$     & A\\
    $a$ (AU)            & $\val{K2-32-circ}{a1}$    & $\val{K2-32-circ}{a2}$     & $\val{K2-32-circ}{a3}$      & A\\
    \Sinc (\Se)         & $\val{K2-32-circ}{Sinc1}$ & $\val{K2-32-circ}{Sinc2}$  & $\val{K2-32-circ}{Sinc3}$   & A\\
    \Teq (K)            & $\val{K2-32-circ}{Teq1}$  & $\val{K2-32-circ}{Teq2}$   & $\val{K2-32-circ}{Teq3}$    & A\\
        \\[-2ex]
    \multicolumn{5}{l}{{\bf Circular RV model (adopted)}} \\
    $K$ (\ms)             & $\val{K2-32-circ}{k1}$           &   $\val{K2-32-circ}{k2}$  & $\val{K2-32-circ}{k3}$ & A \\
    \gam{HIRES} (\ms)     & $\val{K2-32-circ}{gamma_hires}$  &                           &                        & A \\
    \gam{HARPS} (\ms)     & $\val{K2-32-circ}{gamma_harps}$  &                           &                        & A \\
    \gam{PFS} (\ms)       & $\val{K2-32-circ}{gamma_pfs}$    &                           &                        & A \\
    \dvdt (\msyr)         & 0 (fixed)                        &                           &                        & A \\
    \sigjit{HIRES} (\ms)  & $\val{K2-32-circ}{jit_hires}$    &                           &                        & A \\
    \sigjit{HARPS} (\ms)  & $\val{K2-32-circ}{jit_harps}$    &                           &                        & A \\
    \sigjit{PFS} (\ms)    & $\val{K2-32-circ}{jit_pfs}$      &                           &                        & A \\
    \Mp (\Me)             & $\val{K2-32-circ}{Mpsini1}$      & < \val{K2-32-circ}{Mpsiniul2} (95\% conf.)   & $\val{K2-32-circ}{Mpsini3}$ & A \\
    $\rho$ (\gcc)         & $\val{K2-32-circ}{rhop1}$        & < \val{K2-32-circ}{rhopul2} (95\% conf.)     & $\val{K2-32-circ}{rhop3}$   & A \\
        \\[-2ex]
    \multicolumn{5}{l}{{\bf Eccentric RV model}} \\
    $K$ (\ms)             & $\val{K2-32-ecc}{k1}$                  & $\val{K2-32-ecc}{k2}$                      & $\val{K2-32-ecc}{k3}$       & A \\
    $e$                   & < $\val{K2-32-ecc}{eul1}$ (95\% conf.) & Unconstrained                              & Unconstrained               & A \\
    \gam{HIRES} (\ms)     & $\val{K2-32-ecc}{gamma_hires}$         &                                            &                             & A \\
    \gam{HARPS} (\ms)     & $\val{K2-32-ecc}{gamma_harps}$         &                                            &                             & A \\
    \gam{PFS} (\ms)       & $\val{K2-32-ecc}{gamma_pfs}$           &                                            &                             & A \\
    \dvdt (\msyr)         & 0 (fixed)                              &                                            &                             & A \\
    \sigjit{HIRES} (\ms)  & $\val{K2-32-ecc}{jit_hires}$           &                                            &                             & A \\
    \sigjit{HARPS} (\ms)  & $\val{K2-32-ecc}{jit_harps}$           &                                            &                             & A \\
    \sigjit{PFS} (\ms)    & $\val{K2-32-ecc}{jit_pfs}$             &                                            &                             & A \\
    \Mp (\Me)             & $\val{K2-32-ecc}{Mpsini1}$             & < \val{K2-32-ecc}{Mpsiniul2} (95\% conf.)  & $\val{K2-32-ecc}{Mpsini3}$  & A \\
    $\rho$ (\gcc)         & $\val{K2-32-ecc}{rhop1}$               & < \val{K2-32-ecc}{rhopul2} (95\% conf.)    & $\val{K2-32-ecc}{rhop3}$    & A \\
\hline
\end{tabular}
\tablecomments{A: This work; B: \cite{Crossfield16}. Because the 2$\sigma$ confidence interval on \K{c}includes zero, we report upper limits on the mass and density of K2-32.}

\label{tab:k2-32}
\end{table*}
}

\subsection{K2-39}
\label{ssec:k2-39}
\begin{figure}
\includegraphics[width=0.48\textwidth]{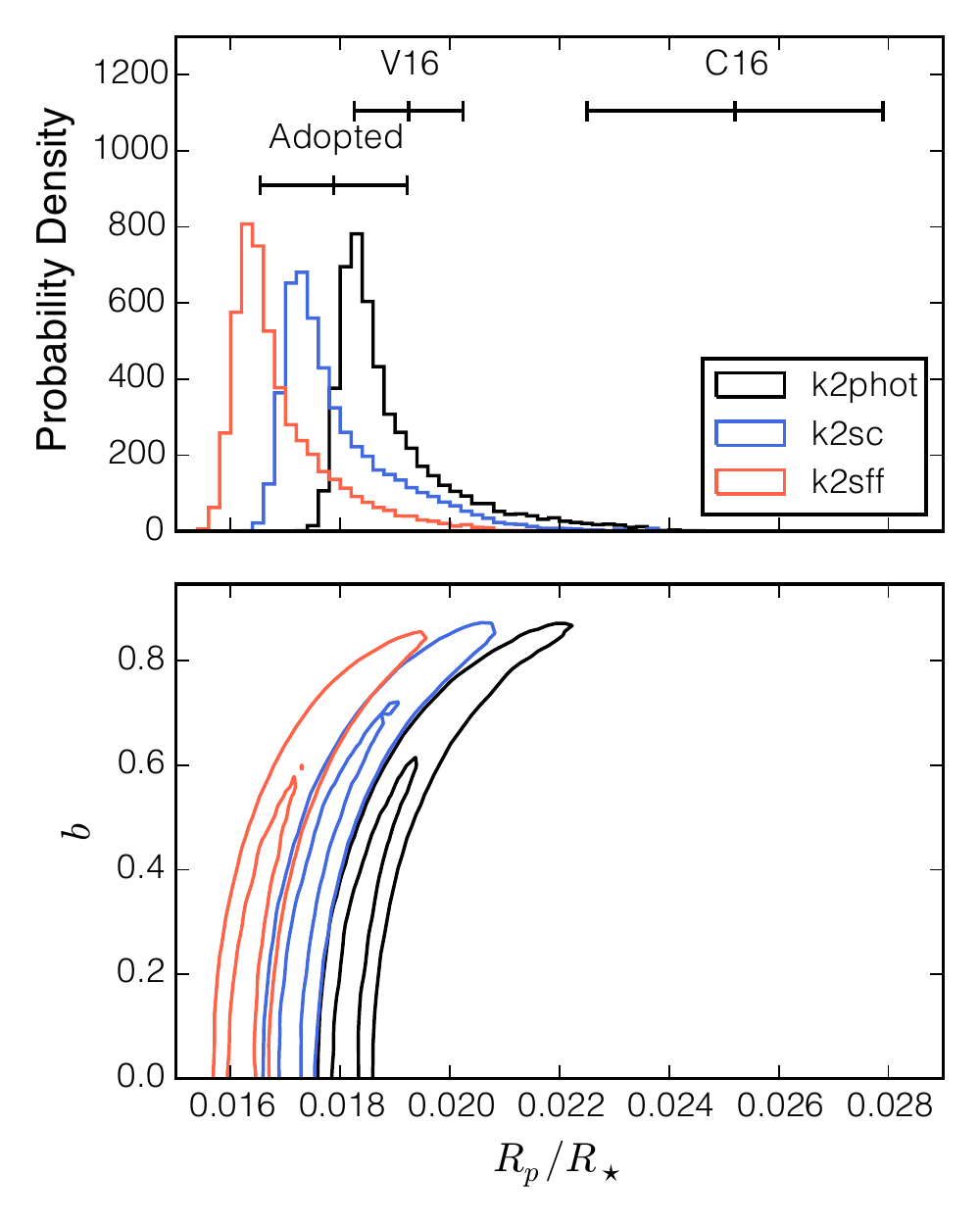}
\caption{Results from MCMC fitting of different photometric reductions of K2-39 light curve, described in Section~\ref{ssec:k2-39}. Top: different values of the \rrat from the literature and this work. Significant disagreement between the \cite{VanEylen16b} and \cite{Crossfield16} values (V16 and C16, respectively), motivated a reanalysis of photometry generated by three independent pipelines: {\tt k2phot}, {\tt k2sc}, and {\tt k2sff}. The histograms show the posterior distributions after an MCMC exploration for different datasets. The different reductions led to posteriors on \Rp and $b$ that differed by $\approx1-3\sigma$, perhaps due to different susceptibilities to correlated noise. Our adopted value combines all three chains to conservatively represent \rrat. Bottom: the 1$\sigma$ and 2$\sigma$ contours for \rrat and the impact parameter, $b$. We see the well-known correlation between \rrat and $b$ for large values of $b$. The \cite{Crossfield16} analysis favored a grazing transit, and thus a larger value of $\rrat$. Given the disagreement between the various reductions, we combined the three MCMC chains to derive a more conservative ``adopted'' value of \Rp/\Rstar.  \label{fig:k2-39-radius}}
\end{figure}

K2-39 hosts a single transiting planet, K2-39b, with $P = 4.60$~d, which was first confirmed by \cite{VanEylen16b}. When fitting the photometry, \cite{VanEylen16b} and \cite{Crossfield16} arrived at different values of the planet-to-star radius ratio, $\Rp/\Rstar$, of $1.93 \pm 0.1$\% and $2.52 \pm 0.27$\%, respectively. The \cite{Crossfield16} solution favored a grazing impact parameter of $b = 1.10^{+0.07}_{-0.09}$ and is responsible for the larger inferred radius ratio. Motivated by this discrepancy we re-examined the K2-39 photometry produced by the {\tt k2phot} pipeline%
\footnote{https://github.com/petigura/k2phot}
and two other publicly available light curves produced by the {\tt k2sff} and {\tt k2sc} pipelines \citep{Vanderburg14,Aigrain15}. For each light curve, we masked out the in-transit points and modeled the out-of-transit photometry with a Gaussian Process \citep{Rasmussen05} using a squared-exponential kernel with a correlation length of 2 days. We then performed a standard MCMC exploration of the likelihood surface using the {\tt batman} \citep{Kreidberg15} and {\tt emcee} Python packages to map out parameter uncertainties and covariances. Figure~\ref{fig:k2-39-radius} summarizes the results of these different fits. The {\tt k2phot}, {\tt k2sc}, and {\tt k2sff} light curves yielded $\Rp/\Rstar$ of  $1.85_{-0.05}^{+0.15}$\%, $1.76_{-0.06}^{+0.15}$\%, and $1.66_{-0.05}^{+0.12}$\%, respectively. We also note the significant correlation between $\Rp/\Rstar$ and $b$ at large values of $b$, which is responsible for the rather asymmetric uncertainties. Because $\Rp/\Rstar$ differs by 2--3$\sigma$ among the different reductions, we combined the three sets of MCMC chains and adopt $\rrat = \val{K2-39-circ}{rrat1}$\% as an intermediate value of the planet-to-star radius ratio with more conservative errors.

We obtained 42 spectra with HIRES of K2-39 between 2015-08-10 and 2016-08-21. \cite{VanEylen16b} obtained 7 spectra with HARPS, 6 with PFS, and 17 with FIES, which we included in our analysis. In contrast to our K2-27 analysis (Section~\ref{ssec:k2-27}), we included FIES RVs because of the larger number of measurements (17 as opposed to 6 for K2-27) and because they are the only non-HIRES dataset with sufficient time baseline to resolve a long timescale activity signal, which we discuss below.

\begin{figure}
\gridline{\fig{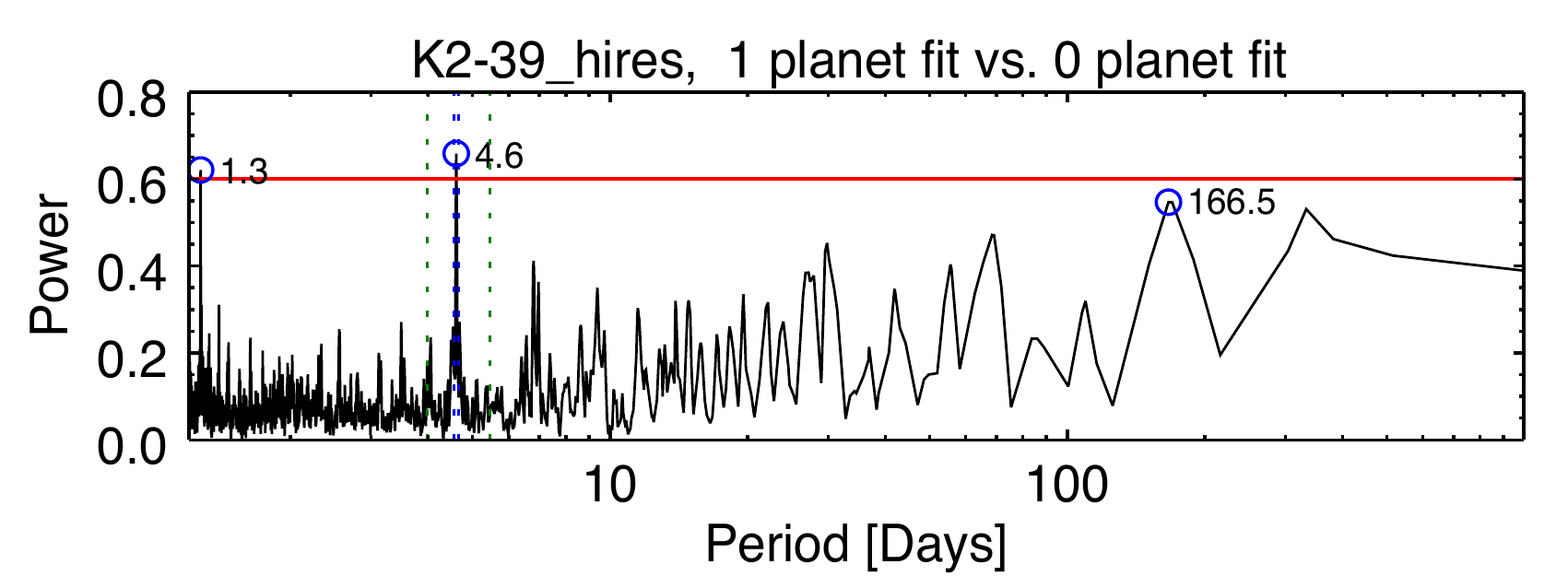}{0.45\textwidth}{}}
\gridline{\fig{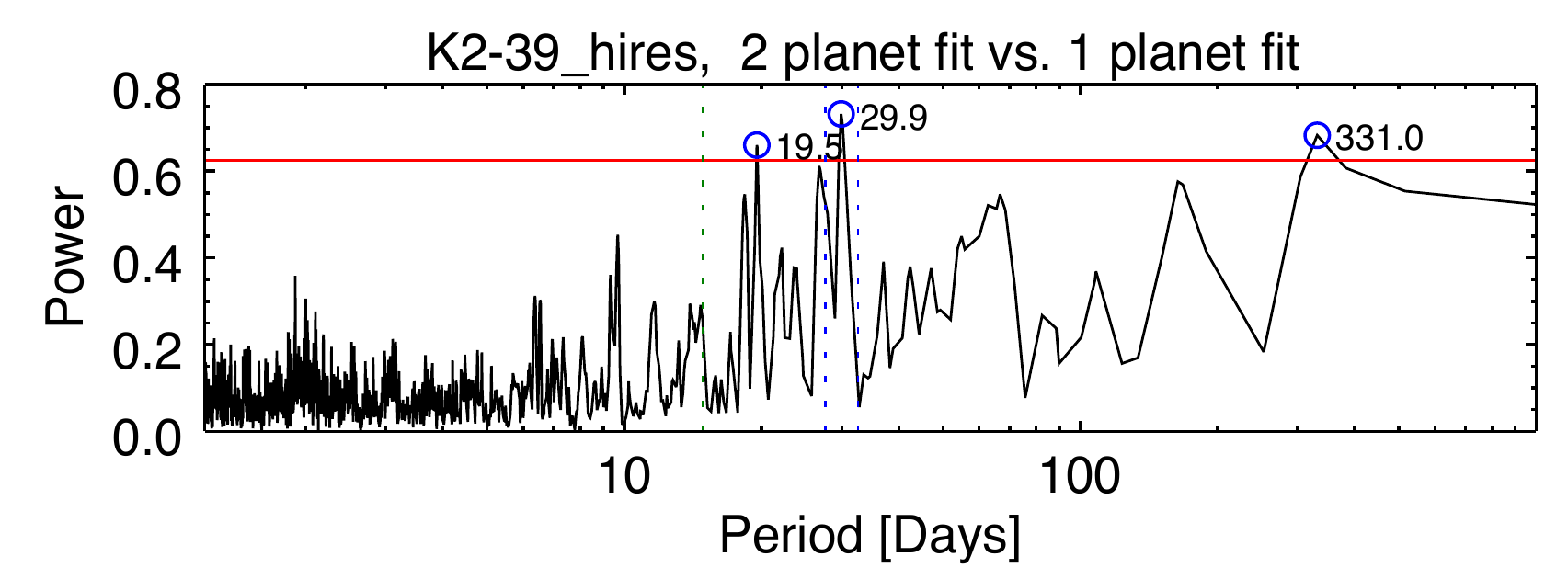}{0.45\textwidth}{}}
\gridline{\fig{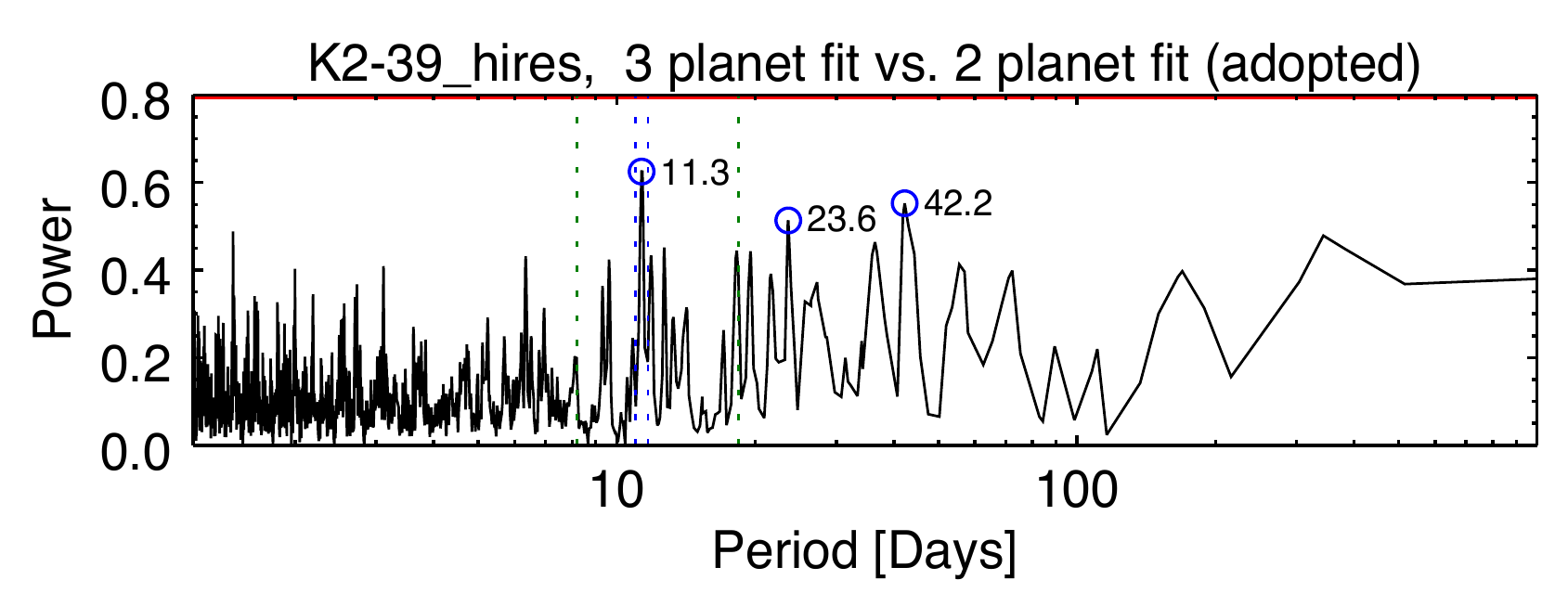}{0.45\textwidth}{}}
\gridline{\fig{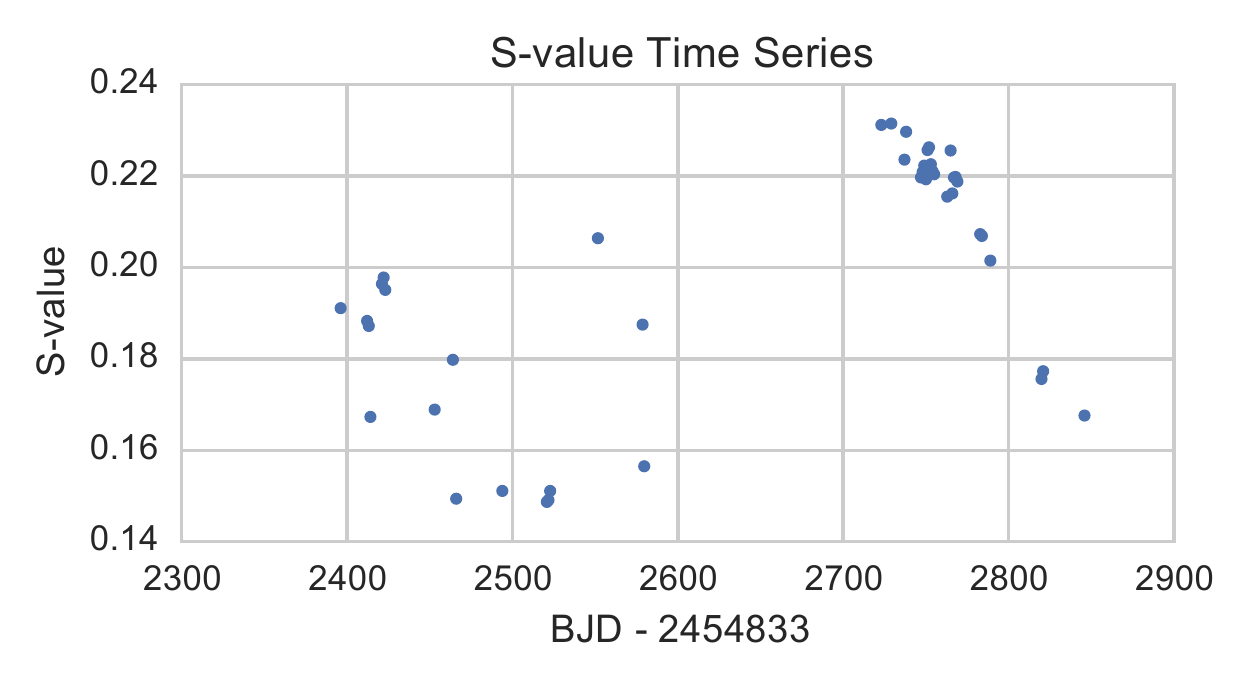}{0.45\textwidth}{}}
\caption{Searches for Keplerian signals in the radial velocity (RV) time series of K2-39 (see Section~\ref{ssec:k2-39}). The top three panels show searches for Keplerian signatures in the HIRES RV time series of K2-39 using a Two-Dimensional Keplerian Lomb-Scargle (2DKLS) periodogram. The bottom panel shows the S-values for K2-39, which traces stellar activity. First panel: The 2DKLS search of the HIRES data reveals a peak at $P = 4.6$~d, which corresponds to K2-39b, the transiting planet. Second panel: Change in $\chi^2$ when comparing a 2-planet fit to a 1-planet fit (power). There are several putative signals that are more significant then the 10\% eFAP threshold plotted in red \citep{Howard16}. However, the $\approx$330~d signal is associated with an activity cycle, seen in the S-values. We fit this activity signal with a Keplerian, and perform a final 2DKLS search for additional planets, shown in the third panel. All other potential signals fall below the 10\% eFAP threshold.}
\label{fig:k2-39-pgram}
\end{figure}

The RVs exhibited large amplitude variability that was not associated with K2-39b that motivated searches for additional non-transiting planets. Figure~\ref{fig:k2-39-pgram} shows three successive Keplerian searches in the K2-39 RVs using a modified version of the Two-Dimensional Keplerian Lomb-Scargle (2DKLS) periodogram \citep{Otoole09,Howard16}. We observed a peak in the periodogram at $P = 4.6$~d, which corresponds to the period of K2-39b. When we measured the change in $\chi^2$ (periodogram power) between a 1-planet fit and a 2-planet fit, then between a 2 planet model and a 3-planet model (lower two panels in Figure \ref{fig:k2-39-pgram}) we found several peaks that fall above the 10\% empirical false alarm threshold (eFAP, \citealt{Howard16}). However, inspection of the S-values  (\citealt{Isaacson10}, see Figure~\ref{fig:k2-39-pgram}), which can correlate with stellar activity that can lead to spurious Doppler signals, showed significant long-term variability that is associated with the RV peak seen at $\approx$330~d and is likely not associated with another planet. We modeled out the activity signal to search for additional non-transiting planets. As a matter of convenience, we modeled the activity signature as a Keplerian and removed it from the timeseries. A subsequent search of the residual RVs revealed no other significant signals.

\begin{figure*}
\plotone{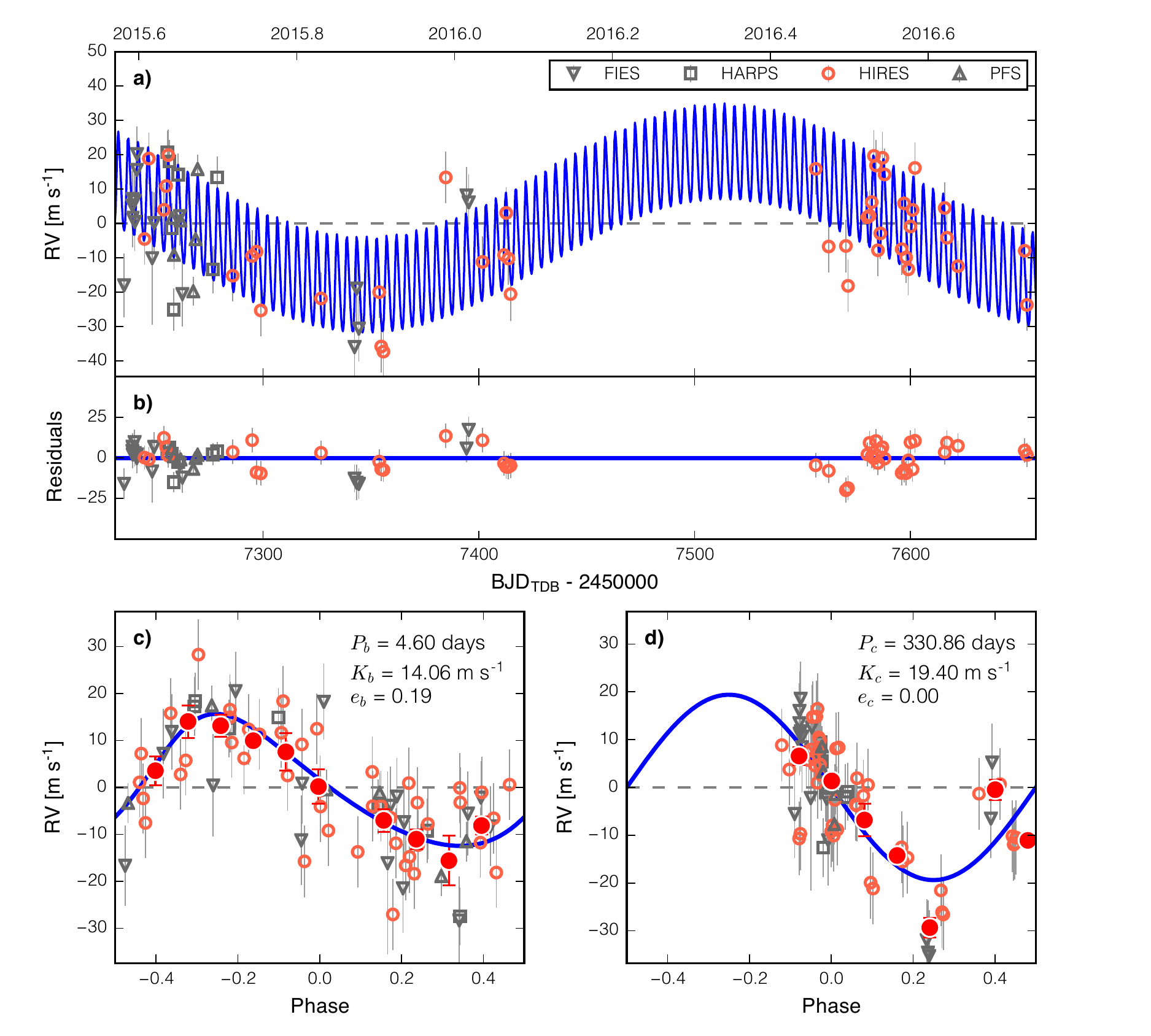}
\caption{Fit to the K2-39 radial velocities (RVs) with with two Keplerians (see Section~\ref{ssec:k2-39}). {\bf a)} Time series of RVs from HIRES along with FIES, HARPS, and PFS published in \cite{VanEylen16b}.  During the fitting we allowed for an arbitrary offset between the four instruments to float as a free parameter. The blue line shows the most probable two-Keplerian model. The long-period Keplerian models out a long period activity cycle, which is also apparent in the S-values. {\bf b)} Residuals to the most probable Keplerian model. Panels {\bf c)} and {\bf d)} show the phase-folded RVs and the most probable Keplerian model with the contributions from the other planets removed. The large red circles show the phase-binned RVs. Note that because the posterior on \e{b} is asymmetric, the eccentricity of the most probable Keperian differs slightly from the value reported in Table~\ref{tab:k2-39} (\e{b} = $\val{K2-39-ecc}{e1}$), which reflects the median of the posterior distribution.\label{fig:k2-39}}
\end{figure*}

We modeled the combined HIRES, HARPS, FIES, and PFS RVs using two Keplerians: one for K2-39b and one as a convenient description of the stellar activity. We first assumed circular orbits and no additional acceleration term, \dvdt. We found that models that included \dvdt were not favored by the BIC and fixed \dvdt = 0~\msyr in subsequent fits. Next, we allowed the eccentricity of K2-39b to float, and found that this model was preferred over the circular model ($\dbic = -7$). We adopted the eccentric model, which is shown in Figure~\ref{fig:k2-39}. The properties of the K2-39 system are listed in Table~\ref{tab:k2-39}. 

For K2-39b, we found a mass of \Mp = $\val{K2-39-ecc}{Mpsini1}$~\Me, which agrees with $\Mp = 50.3^{+9.7}_{-9.4}$~\Me reported by \cite{VanEylen16b} at the 1$\sigma$ level. The additional RVs improved the precision of the mass measurement by roughly a factor of two. Our derived planet radius of $\Rp = \val{K2-39-ecc}{Rp1}$~\Re is substantially smaller than $\Rp = 8.2\pm1.1$~\Re reported by \cite{VanEylen16b}. This is largely due to the smaller stellar radius (see Section~\ref{sec:stellar}). Our adopted value of $\rrat = \val{K2-39-circ}{rrat1}$\% is also smaller than the \cite{VanEylen16b} value of \rrat = $1.93 \pm 0.1$\%. While the difference in \rrat is not as large in a fractional sense as the difference in \Rstar, it also contributes to a smaller derived \Rp. Thus, our derived density of $\rho = \val{K2-39-ecc}{rhop1}$~\gcc, is significantly larger than $\rho = 0.50^{+0.29}_{-0.17}$~\gcc, reported by \cite{VanEylen16b}.

{\renewcommand{\arraystretch}{0.9}
\begin{table*}
\centering
\caption{K2-39 planet parameters}
\begin{tabular}{lrrr}
\hline
\hline
                          & {\rm Value }  &  & {\rm Ref.}  \\
\hline
    \multicolumn{4}{l}{{\bf Stellar parameters}} \\
    Identifiers         & EPIC-206247743              &  & \\
                        & TYC-5811-835-1              &  & \\
    \teff (K)           & $\val{K2-39-star}{teff}$    &  & A \\
    \logg (dex)         & $\val{K2-39-star}{logg}$    &  & A \\
    \fe (dex)           & $\val{K2-39-star}{fe}$      &  & A \\
    \vsini (\kms)       & $\val{K2-39-star}{vsini}$   &  & A \\
    \Mstar (\Msun)      & $\val{K2-39-star}{mass}$    &  & A \\
    \Rstar (\Rsun)      & $\val{K2-39-star}{radius}$  &  & A \\
    age (Gyr)           & $\val{K2-39-star}{agegyr}$  &  & A \\
    Apparent $V$ (mag)  & $\val{K2-39-star}{vmag}$    &  & A \\
    Spec. parallax      & 

    \\[-2ex]
    {} & {\bf planet b} & {\bf activity} &  \\ 
    \multicolumn{4}{l}{{\bf Transit model}} \\
    $P$ (days)            & $\val{K2-39-ecc}{P1}$     &   & B\\
    $T_0$ (BJD-2454833)   & $\val{K2-39-ecc}{T01}$    &   & B\\
    \rrat (\%)            & $\val{K2-39-circ}{rrat1}$ &   & A\\
    \Rp (\Re)             & $\val{K2-39-ecc}{Rp1}$    &   & A\\
    $a$ (AU)              & $\val{K2-39-ecc}{a1}$     &   & A\\
    \Sinc (\Se)           & $\val{K2-39-ecc}{Sinc1}$  &   & A\\
    \Teq (K)              & $\val{K2-39-ecc}{Teq1}$   &   & A\\
    \\[-2ex]
    \multicolumn{4}{l}{{\bf Circular RV model}} \\
    $P$ (days)           & fixed                           &  $\val{K2-39-circ}{per2}$ & B\\
    $T_0$ (BJD)          & fixed                           &  $\val{K2-39-circ}{tc2}$  & B\\
    $K$ (\ms)            & $\val{K2-39-circ}{k1}$          &  $\val{K2-39-circ}{k2}$   & A \\
    \gam{HIRES} (\ms)    & $\val{K2-39-circ}{gamma_hires}$ &                           & A \\
    \gam{HARPS} (\ms)    & $\val{K2-39-circ}{gamma_harps}$ &                           & A \\
    \gam{PFS} (\ms)      & $\val{K2-39-circ}{gamma_pfs}$   &                           & A \\
    \gam{FIES} (\ms)     & $\val{K2-39-circ}{gamma_fies}$  &                           & A \\
    \dvdt (\msyr)        & 0 (fixed)                       &                           & A \\
    \sigjit{HIRES} (\ms) & $\val{K2-39-circ}{jit_hires}$   &                           & A \\
    \sigjit{HARPS} (\ms) & $\val{K2-39-circ}{jit_harps}$   &                           & A \\
    \sigjit{PFS} (\ms)   & $\val{K2-39-circ}{jit_pfs}$     &                           & A \\
    \sigjit{FIES} (\ms)  & $\val{K2-39-circ}{jit_fies}$    &                           & A \\
    \Mp (\Me)            & $\val{K2-39-circ}{Mpsini1}$     &  --                       & A \\
    $\rho$ (\gcc)        & $\val{K2-39-circ}{rhop1}$       &  --                       & A \\
    \\[-2ex]
    \multicolumn{4}{l}{{\bf Eccentric RV model (adopted)}} \\
    $P$ (days)           & fixed                          & $\val{K2-39-ecc}{per2}$ & B\\
    $T_0$ (BJD)          & fixed                          & $\val{K2-39-ecc}{tc2}$  & B\\
    $K$ (\ms)            & $\val{K2-39-ecc}{k1}$          & $\val{K2-39-ecc}{k2}$   & A \\
    $e$                  & $\val{K2-39-ecc}{e1}$          & $\val{K2-39-ecc}{e2}$   & A \\
    \gam{HIRES} (\ms)    & $\val{K2-39-ecc}{gamma_hires}$ &                         & A \\
    \gam{HARPS} (\ms)    & $\val{K2-39-ecc}{gamma_harps}$ &                         & A \\
    \gam{PFS} (\ms)      & $\val{K2-39-ecc}{gamma_pfs}$   &                         & A \\
    \gam{FIES} (\ms)     & $\val{K2-39-ecc}{gamma_fies}$  &                         & A \\
    \dvdt (\msyr)        & 0 (fixed)                      &                         & A \\
    \sigjit{HIRES} (\ms) & $\val{K2-39-ecc}{jit_hires}$   &                         & A \\
    \sigjit{HARPS} (\ms) & $\val{K2-39-ecc}{jit_harps}$   &                         & A \\
    \sigjit{PFS} (\ms)   & $\val{K2-39-ecc}{jit_pfs}$     &                         & A \\
    \sigjit{FIES} (\ms)  & $\val{K2-39-ecc}{jit_fies}$    &                         & A \\
    \Mp (\Me)            & $\val{K2-39-ecc}{Mpsini1}$     & --                      & A \\
    $\rho$ (\gcc)        & $\val{K2-39-ecc}{rhop1}$       & --                      & A \\
\hline
\end{tabular}
\tablecomments{A: This work; B: \cite{Crossfield16}. We model a large amplitude ($\approx 20$\ms) stellar activity signal by introducing an additional circular Keplerian.}
\label{tab:k2-39}
\end{table*}
}

\subsection{K2-108}
\label{ssec:epic2117}
\begin{figure*}
\plotone{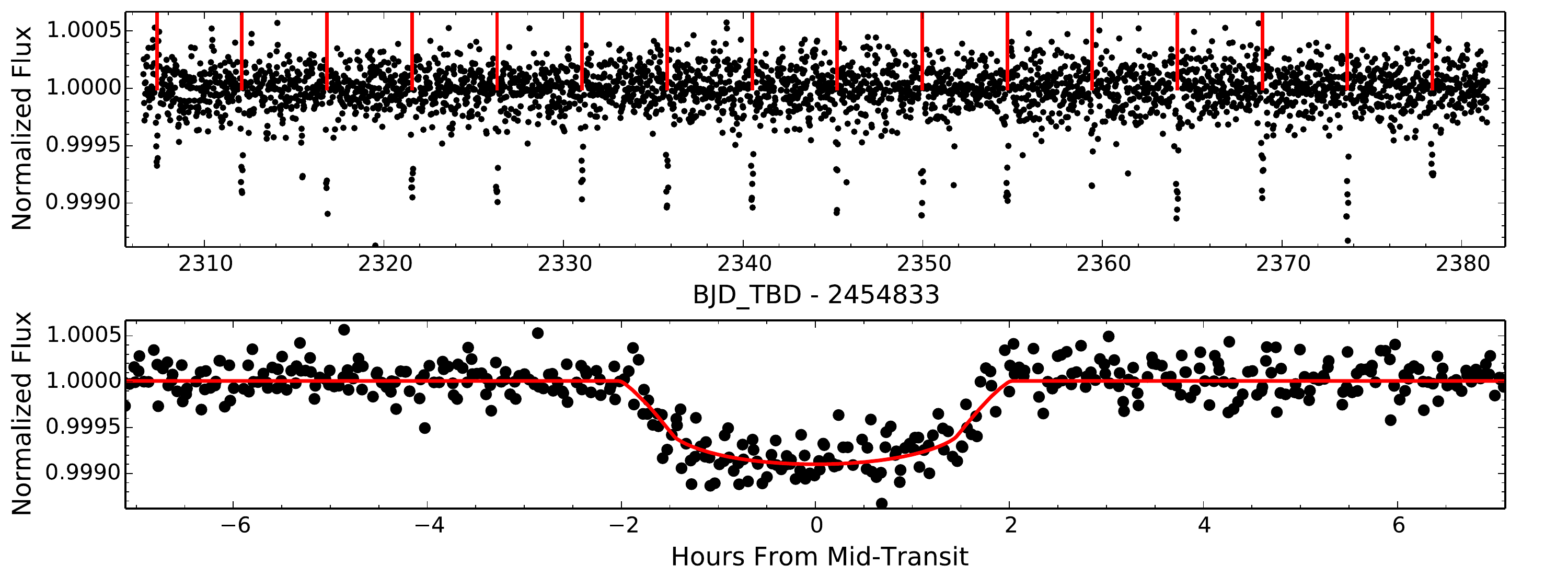}
\caption{Top: \ktwo light curve of K2-108 showing with the transits of K2-108b labeled with red ticks. Bottom: photometry phase-folded on the transit ephemeris.\label{fig:epic2117-lightcurve}}
\end{figure*}

K2-108, listed as EPIC-211736671 in the Ecliptic Planet Input Catalog \citep{Huber16}, is a $V = 12.3$~mag star observed during \ktwo Campaign 5. We identified K2-108 as a likely planet according to our team's standard methodology, described in detail in \cite{Crossfield16}. In brief, we identified a set of transits having $P = 4.73$~d and elevated K2-108 to the status of ``planet candidate.'' We fit the light curve according to standard procedures and show the best-fitting model light curve in Figure~\ref{fig:epic2117-lightcurve}. Follow-up spectroscopic observations revealed that K2-108 is a metal-rich ($\fe = \val{epic2117-star}{fe}$~dex), slightly-evolved G star having a radius of \Rstar = $\val{epic2117-star}{radius}$~\Rsun. The spectroscopically-determined stellar parameters along with the results from our light curve fitting are listed in Table~\ref{tab:epic2117}. 

\begin{figure*}
\plotone{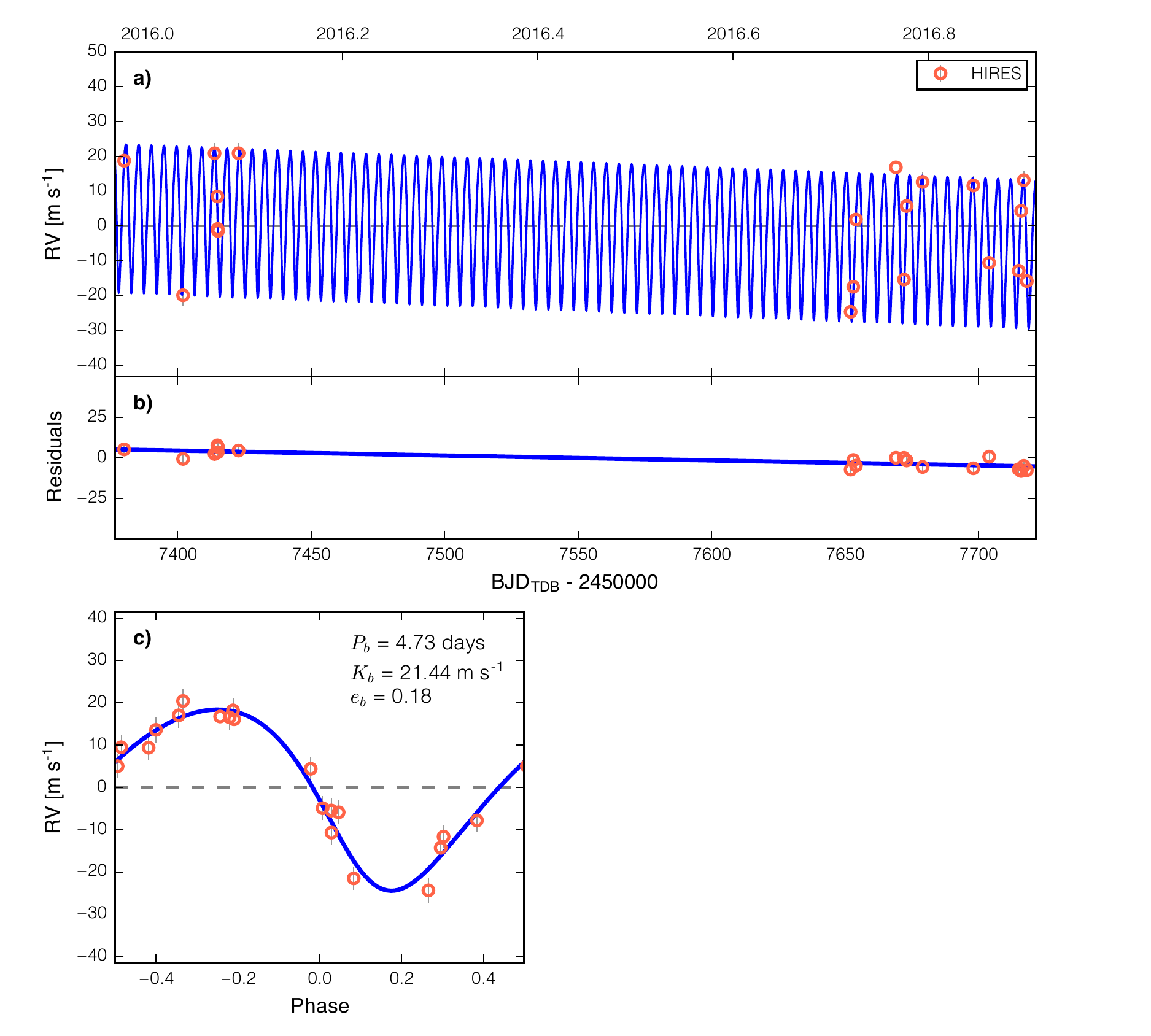}
\caption{Single Keplerian model to the K2-108 radial velocities (RVs), allowing for eccentricity (see Section~\ref{ssec:epic2117}). {\bf a)} Time series of RVs from HIRES. The blue line shows the most probable Keplerian model. {\bf b)} Residuals to the most probable Keplerian model. {\bf c)} The phase-folded RVs and the most probable Keplerian model.\label{fig:epic2117}}.
\end{figure*}

We obtained 20 spectra with HIRES of K2-108 between 2015-12-23 and 2016-11-25. We first considered circular models with no acceleration term, \dvdt. We found that eccentric models with non-zero \dvdt (shown in Figure~\ref{fig:epic2117}) were favored over circular models ($\dbic = -25$). The parameters are summarized in Table~\ref{tab:epic2117}. At $\val{epic2117-ecc}{Mpsini1}$~\Me, K2-108b is remarkably massive for a $\val{epic2117-ecc}{Rp1}$~\Re planet, implying a large heavy element component.

While our RV analysis verified the planetary nature of the transiting object, we assessed the possibility of additional stellar companions in the photometric aperture that could appreciably dilute the observed transit, resulting in an incorrect derived planetary radius.  From the light curve fits, the planet-to-star radius ratio is $\Rp/\Rstar = 2.82\pm0.17\%$. For an additional star to significantly alter the observed transit depth requires a flux ratio, $F_2/F_1 \approx \sigma\left(\left(\Rp/\Rstar\right)^2\right) / \left(\Rp/\Rstar\right)^2 \approx 0.10$ or $\Delta V \approx 2.5$~mag \citep{Ciardi15}.

A search for secondary spectral lines in the HIRES template spectrum \citep{kolbl:2015} revealed no additional stellar companions having $\Delta V \lesssim 5$~mag and $\Delta \mathrm{RV} \gtrsim 15$~\kms, corresponding to a physical separation of $\lesssim 1$~\AU. 

\begin{figure}
\centering
\includegraphics[width=0.5\textwidth]{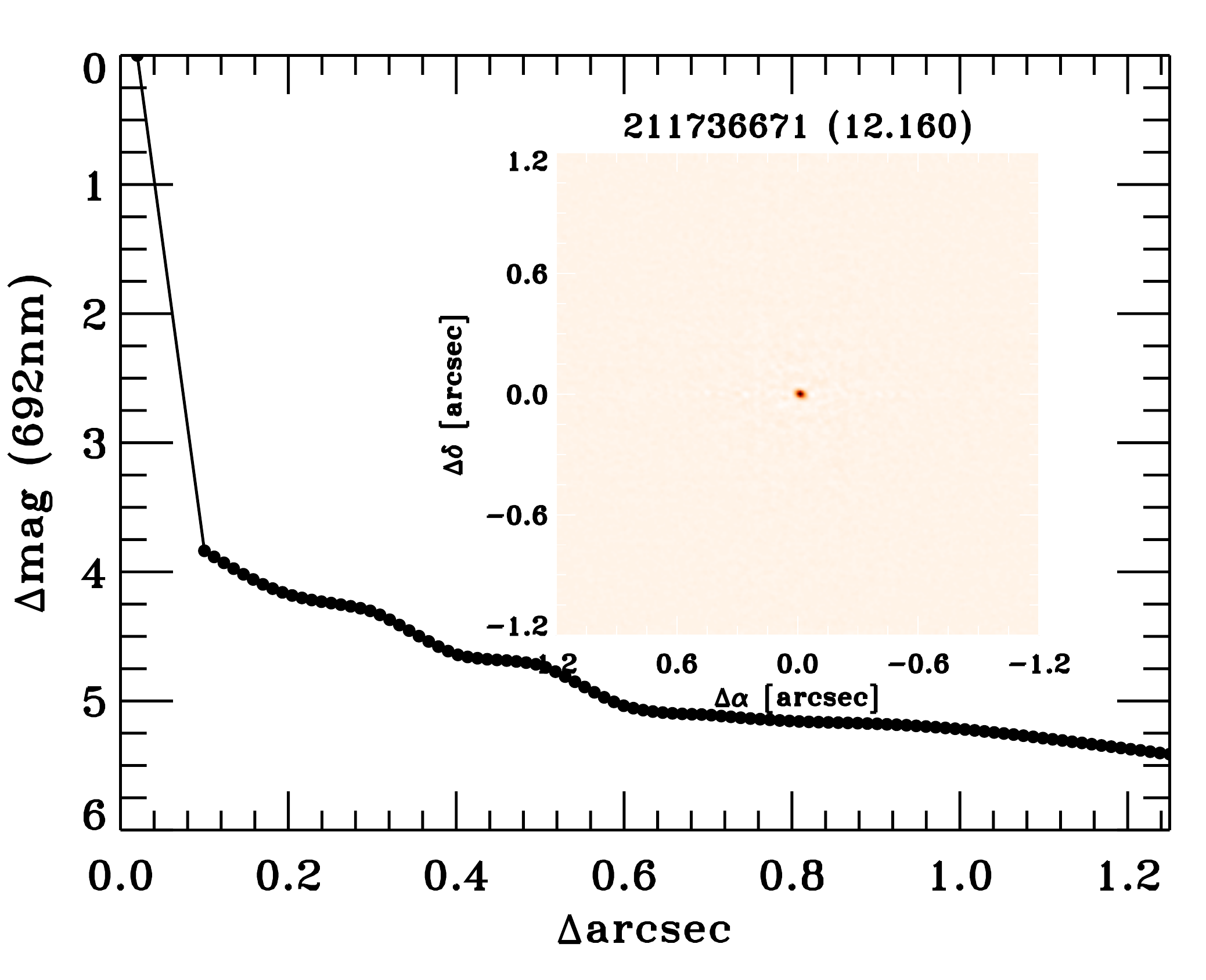}
\caption{Contrast curve of K2-108 (EPIC-211736671) taken with a narrow-band filter centered at 692~nm using the DSSI camera \citep{Horch12} on the Gemini-N 8m telescope. The inset image shows the $2 \times 2$ arcsec region centered on K2-108 reconstructed from the speckle image sequence. Putative companions with contrasts of $<$2.5~mag which could alter the inferred planet size are ruled out for separations $> 80$~mas. Additional DSSI observations at 880~nm and Keck/NIRC2 observations at $K$-band also reveal no additional companions to K2-108. See Section \ref{ssec:epic2117} for further details.\label{fig:epic2117-dssi}} 
\end{figure}

We obtained high-resolution speckle imaging at 692~nm with the Differential Speckle Survey Instrument (DSSI; \citealt{Horch12}), a visitor instrument used at the Gemini-North 8.1-m  telescope on 2016-01-13. The observations revealed no additional companions with $\Delta V$ < 2.5~mag in the \ktwo photometric aperture down to separations of 80~mas, or $\approx32$~AU in projected separation at the distance of K2-108.%
\footnote{
$d \approx 400$~pc computed according to the same technique used for K2-39 (see Section~\ref{sec:stellar}).  
}
Additional high-resolution imaging with DSSI at 880~nm and Keck/NIRC2 at K-band also revealed no additional companions. Of the three high-resolution images, the DSSI image at 692~nm (shown in Figure~\ref{fig:epic2117-dssi}), provides the tightest contrast curve in the Kepler bandpass. All imaging datasets are available on the Exoplanet Follow-up Observing Program (ExoFOP) website.%
\footnote{\url{https://exofop.ipac.caltech.edu/k2/}}

A stellar companion having $a \approx 1$--$40$~\AU would have evaded the aforementioned follow-up observations. However, such an object would induce a reflex acceleration on K2-108, which would be easily detectable by our RVs:
\[
\dvdt \approx \frac{G \Mstar}{a^2} \approx 120\, \msyr \left(\frac{\Mstar}{\Msun}\right) \left(\frac{a}{40\, \AU}\right)^{-2}.
\]
Over the $\sim$1 year observation baseline, we see only weak evidence for a long-term acceleration ($\dvdt = \val{epic2117-ecc}{dvdt}$~\msyr). To be consistent with the observed \dvdt, a companion at $\approx40~\AU$ would be $\lesssim 0.1~\Msun$ and much too faint to alter the derived planet radius. At smaller separations, the limits on the mass of a putative companion grow more stringent.

{\renewcommand{\arraystretch}{0.9}
\begin{table}
\centering
\caption{System parameters of K2-108 }
\begin{tabular}{lrr}
\hline
\hline
   {\rm Value }   & {\rm Ref.}  \\
\hline
    \multicolumn{3}{l}{{\bf Stellar parameters}} \\
    Identifier                  & EPIC-211736671                & \\
    \teff (K)                   & $\val{epic2117-star}{teff}$   & A \\
    \logg (dex)                 & $\val{epic2117-star}{logg}$   & A \\
    \fe (dex)                   & $\val{epic2117-star}{fe}$     & A \\
    \vsini (\kms)               & $\val{epic2117-star}{vsini}$  & A \\
    \Mstar (\Msun)              & $\val{epic2117-star}{mass}$   & A \\
    \Rstar (\Rsun)              & $\val{epic2117-star}{radius}$ & A \\
    age (Gyr)                   & $\val{epic2117-star}{agegyr}$ & A \\
    Apparent $V$ (mag)          & $\val{epic2117-star}{vmag}$   & A \\
    \\[-2ex]
    {} & {\bf planet b} &  \\ 
    \multicolumn{3}{l}{{\bf Transit model}} \\
    $P$ (days)           & $\val{epic2117-ecc}{P1}$      & A\\
    $T_0$ (BJD-2454833)  & $\val{epic2117-ecc}{T01}$     & A\\
    \Rp (\Re)            & $\val{epic2117-ecc}{Rp1}$     & A\\
        $a$ (AU)         & $\val{epic2117-ecc}{a1}$      & A\\
        \Sinc (\Se)      & $\val{epic2117-ecc}{Sinc1}$   & A\\
        \Teq (K)         & $\val{epic2117-ecc}{Teq1}$    & A\\
    \\[-2ex]
    \multicolumn{3}{l}{{\bf Circular RV model}} \\
    $K$ (\ms)            & $\val{epic2117-circ}{k1}$           & A \\
    \gam{HIRES} (\ms)    & $\val{epic2117-circ}{gamma_hires}$  & A \\
    \dvdt (\msyr)        & 0 (fixed)                           & A \\
    \sigjit{HIRES} (\ms) & $\val{epic2117-circ}{jit_hires}$    & A \\
    \Mp (\Me)            & $\val{epic2117-circ}{Mpsini1}$      & A \\
    $\rho$ (\gcc)        & $\val{epic2117-circ}{rhop1}$        & A \\
    \\[-2ex]
    \multicolumn{3}{l}{{\bf Eccentric RV model (adopted)}} \\
    $K$ (\ms)            & $\val{epic2117-ecc}{k1}$           & A \\
    $e$                  & $\val{epic2117-ecc}{e1}$           & A \\
    \gam{HIRES} (\ms)    & $\val{epic2117-ecc}{gamma_hires}$  & A \\
    \dvdt (\msyr)        & $\val{epic2117-ecc}{dvdt}$         & A \\
    \sigjit{HIRES} (\ms) & $\val{epic2117-ecc}{jit_hires}$    & A \\
    \Mp (\Me)            & $\val{epic2117-ecc}{Mpsini1}$      & A \\
    $\rho$ (\gcc)        & $\val{epic2117-ecc}{rhop1}$        & A \\
\hline
\end{tabular}
\tablecomments{A: This work}
\label{tab:epic2117}
\end{table}
}

\section{Discussion}
\label{sec:discussion}
\subsection{Mass and radius}
\label{ssec:mass-and-radius}
Here, we put the four sub-Saturns presented in this paper in the context of other planets in their size class. Starting with the database of exoplanet properties hosted at the NASA Exoplanet Archive (NEA; \citealt{Akeson13}), we constructed a list of sub-Saturns from the literature that have densities measured to 50\% or better. We supplemented the list with additional measurements that have yet to be ingested into the NEA and removed a few planets with unreliable measurements. Including the measurements from this work, we found 19 sub-Saturns that passed our quality cuts and are listed in Table~\ref{tab:subsaturns}.

\begin{figure*}
\gridline{\fig{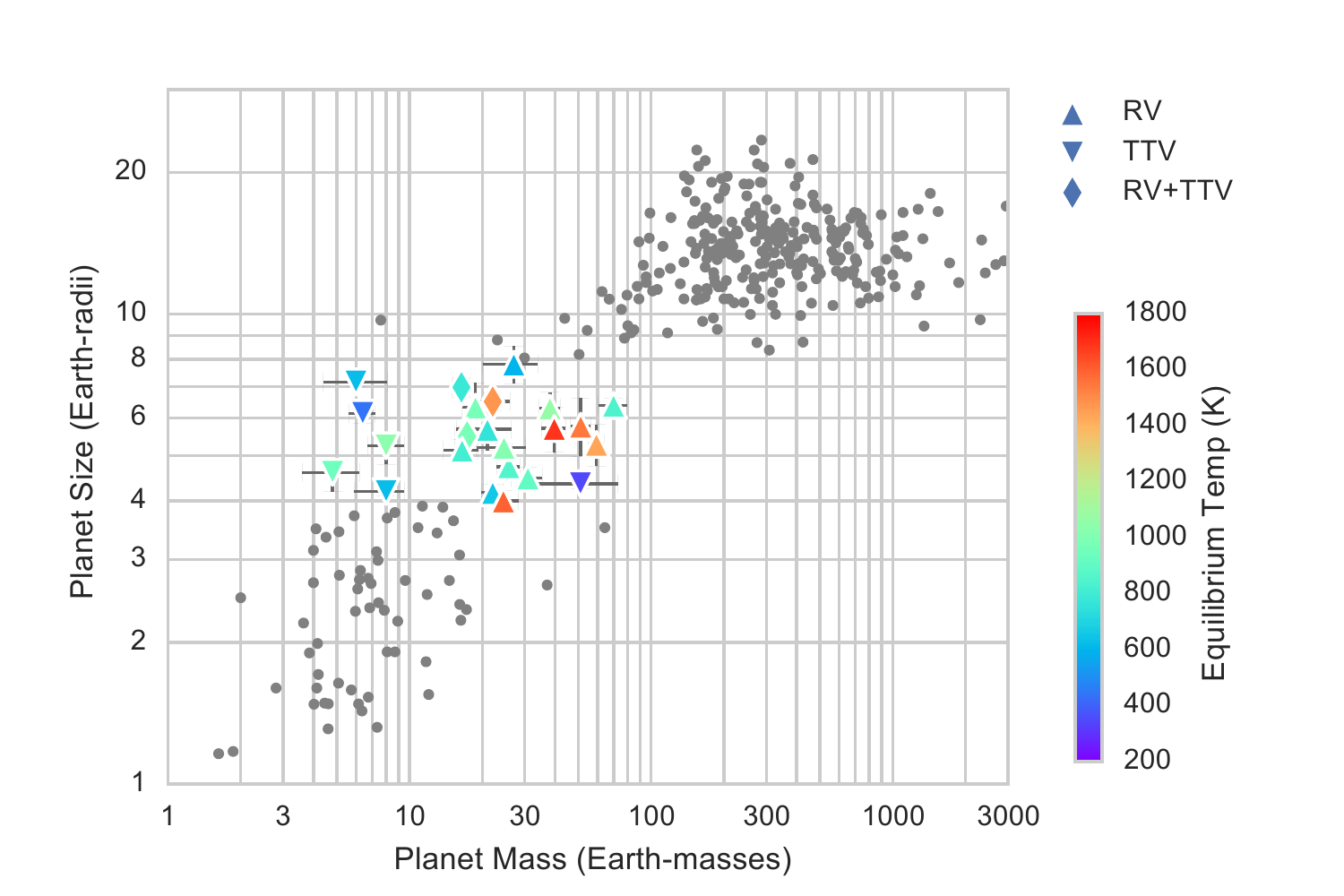}{0.8\textwidth}{}}
\gridline{\fig{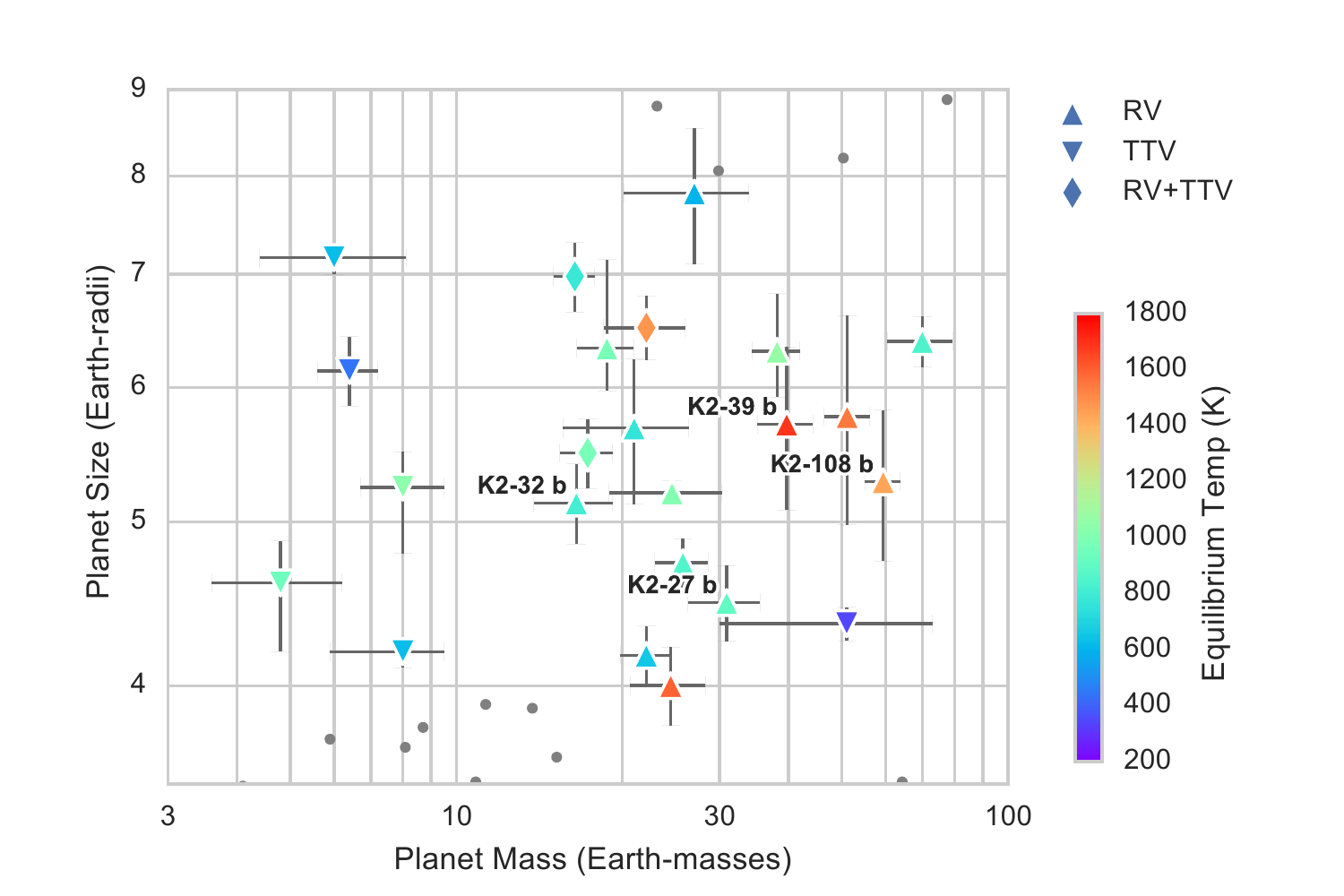}{0.8\textwidth}{}}
\caption{Top: Masses and radii of sub-Saturns having densities measured to 50\% or better. Planets of different size classes with comparable density precision are shown as gray points for context. For the sub-Satruns, symbol colors correspond to the blackbody equilibrium temperature. The symbol shapes correspond to the method by which planet masses were measured: radial velocities (RVs), transit-timing variations (TTVs), or a combined analysis (RVs+TTVs). For sub-Saturns, we note almost no correlation between planet mass and planet size. Bottom: a zoomed in view of the top panel, focusing on sub-Saturns. See Section~\ref{ssec:mass-and-radius} for additional details. \label{fig:mp-rp}}
\end{figure*}

\begin{figure*}
\gridline{\fig{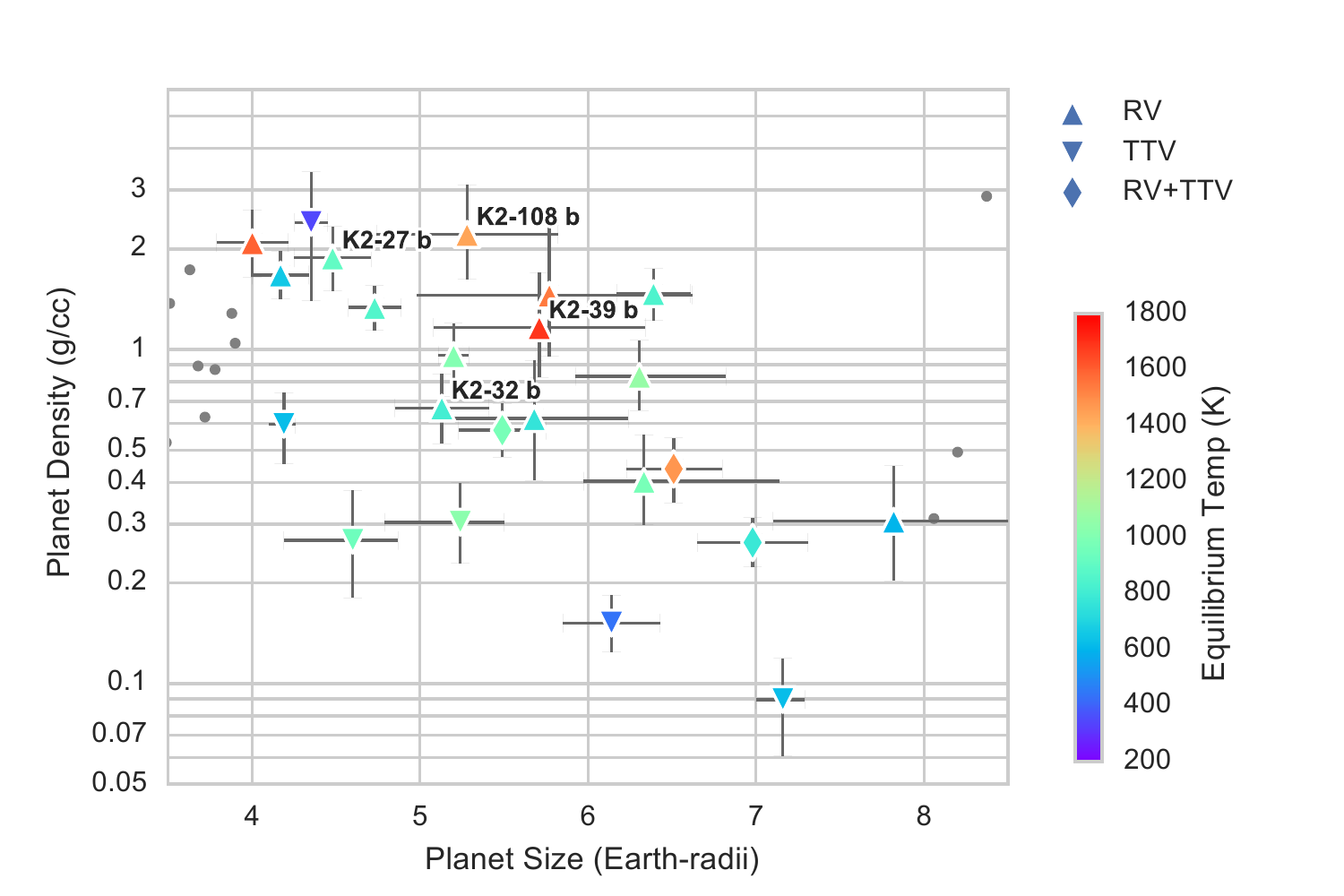}{0.8\textwidth}{}}
\caption{Same as Figure~\ref{fig:mp-rp} but showing mean planet density as a function of planet size. For planets of a given size, there is a diversity of densities due to the diversity in planet mass. While the hottest planets seem to have high densities for their size, cool planets span a wide range of density.\label{fig:rp-rhop}}
\end{figure*}

In Figure~\ref{fig:mp-rp}, we show the measured masses and sizes of these planets, highlighting the measurements from this work. Remarkably, for sub-Saturns, there is little correlation between a planet's mass and size. These planets have a nearly uniform distribution of mass from $\Mp \approx 6-60$~\Me. We note that the hottest planets tend to have higher masses, while cool planets have both high and low masses. Figure~\ref{fig:rp-rhop} shows planet mass and radius, cast in terms of planet size and planet density. The decreasing densities toward larger sizes can be understood simply in terms of larger radii, given there is no strong trend of larger planet masses with larger planet size.

\subsection{Planetary Envelope Fraction}
\label{ssec:cmf}
One advantage of sub-Saturns is that in this size range, both heavy elements and low density gaseous envelopes contribute significantly to the total planet mass. A consequence is, to first order, sub-Saturns can be approximated as two-component planets consisting of a rocky heavy element core, surrounded by an envelope of H/He \citep{Lopez14,Petigura16}. Here we use the results of \cite{Lopez14}, which simulated the internal structure and thermal evolution of planets with solar-composition H/He envelopes atop Earth-composition cores. These simulations were run over a wide swath of parameter space for planets with different masses, \Mp, different envelope fractions, $\fenv = \Menv / \Mp$, and on orbits receiving different levels of incident stellar irradiation, \Sinc. The model planets were allowed to evolve over time and \cite{Lopez14} noted the planet radii at specified intervals. The result of these simulations is a four-dimensional grid of planet radius, \Rp, sampled at various combinations of (\Mp, \fenv, \Sinc, age). 

We used this grid to solve for the values of planet \fenv that are consistent with the observed (\Mp, \Rp, \Sinc, and age). We drew \Mp, \Rp, and \Sinc measured posterior distributions and interpolated the model grid in order to derive \fenv. We assumed a uniform age of 5~Gyr. With the exception of a few stars analyzed with asteroseismology, most of the system ages are derived from isochrone-fitting and are thus uncertain at the $\approx$2--3~Gyr level. Fortunately, the derived \fenv fraction is not sensitive to the adopted system age. Adopting a uniform age of 2~Gyr resulted in typical change in the derived \fenv of $\approx2\%$. The resulting values of \fenv are listed in Table~\ref{tab:subsaturns}.

We have made several approximations when computing \fenv. We have ignored the possibility of water or other volatile ices contributing significantly to heavy element cores of these planets. \citet{Lopez14}, however, showed that including ices in the core, does not significantly alter the radius-composition relationship for planets in this size range. Because the H/He envelope represents most of the planet volume, our inferred \fenv does not depend sensitively on the detailed composition of the planet core. We have also assumed that all the heavy elements are concentrated in the planets' cores rather than being distributed throughout the envelope. 

\subsection{Role of Photo-Evaporation among Sub-Saturns}
If photo-evaporation plays a dominant role in sculpting the gaseous envelopes of sub-Saturns, one might expect the envelope fraction, \fenv, to correlate with the energy it receives from its host star. Figure~\ref{fig:teq} shows the total planet mass (\Mp), and envelope fraction (\fenv) as a function of the blackbody equilibrium temperature. As a matter of convenience, we include core mass (\Mcore) and envelope mass (\Menv) which can be trivially computed from \Mp and \fenv. We do not observe a strong one-to-one correlation envelope fraction and equilibrium temperature. We do, however, observe that cool planets span a large range of $\fenv \approx 10-50\%$ while hot planets span a more narrow range of $\fenv \approx 10-20\%$. Perhaps photo-evaporation excludes planets from occupying certain domains in the \fenv--\Teq plane.

\begin{figure*}
\epsscale{0.9}
\plotone{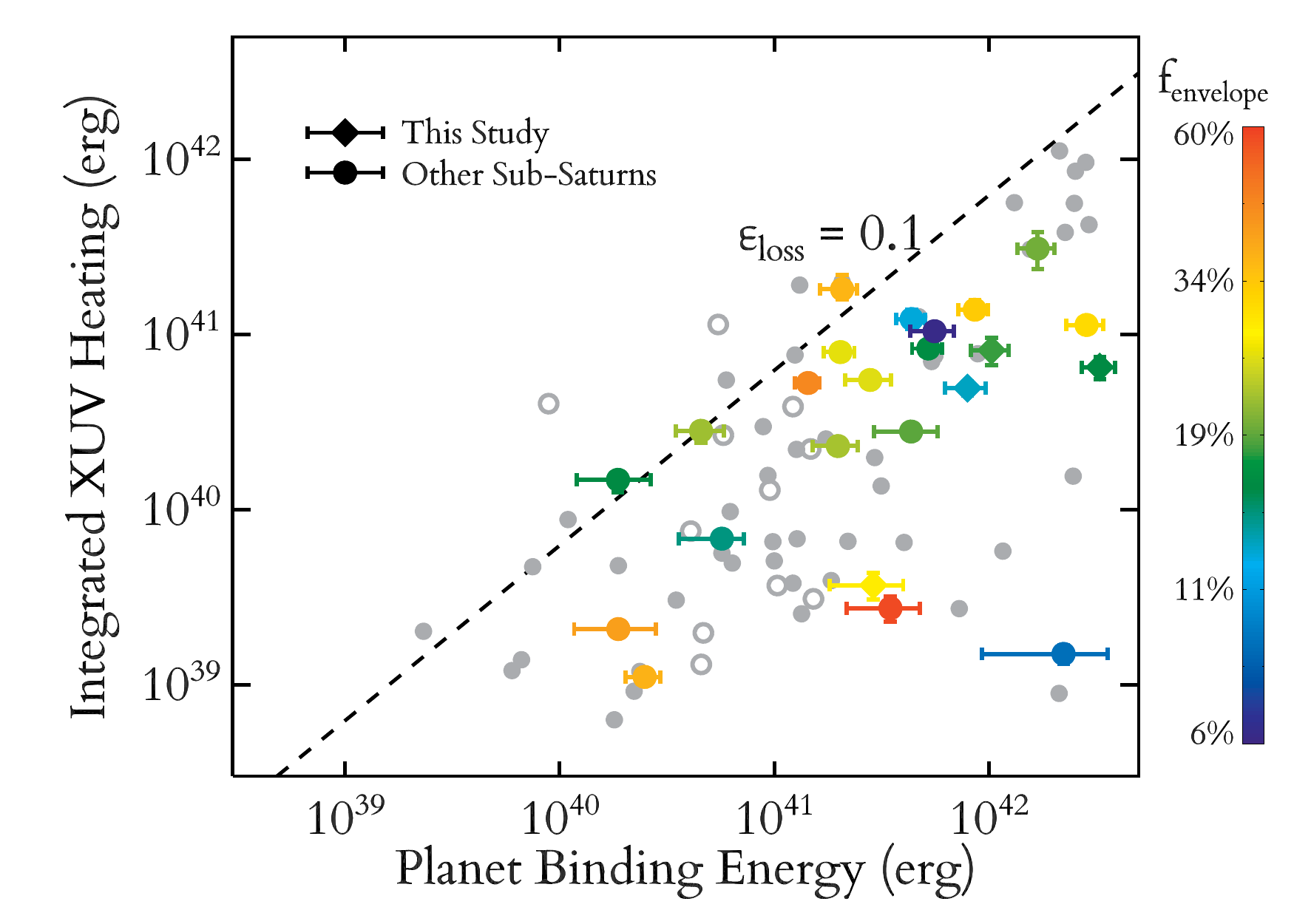}
\caption{The total lifetime XUV heating planets receive at their orbit vs. their current binding energy. Sub-Saturns are color-coded by their current gaseous envelope fraction, while all other planets <100 \Me are shown in grey. The dashed line meanwhile shows the predicted photo-evaporation threshold from \citet{lopez:2012}. This indicates that only a few of the sub-Saturns in this sample have likely been strongly affected by photo-evaporation.\label{fig:massloss}}
\end{figure*}

To further explore the possible role of photo-evaporation, we considered two quantities that are more directly related to a planets susceptibility to photo-evaporation: XUV heating and planet binding energy. XUV heating is the total lifetime-integrated XUV flux incident at a planet's orbit multiplied by the planet's current cross section. Figure~\ref{fig:massloss}, which is updated from \citep{Lopez14}, shows the planet XUV heating (in ergs) vs. planet binding energy (in ergs). This sort of diagram has been used as supporting evidence for the role of photo-evaporation in sculpting the envelopes of highly-irradiated sub-Neptunes \citep{Lecavelier07,lopez:2012,Owen13}. The dashed line shows the envelope survival threshold predicted by photo-evaporation and thermal evolution models in \citet{lopez:2012}, and the absence of planets with gaseous envelopes above this line suggests that planets near this threshold have experienced photo-evaporation. While a few sub-Saturns lie close to this threshold, most lie well below it, indicating that they are sufficiently massive or are on wide enough orbits to be immune to significant photo-evaporation. 

\subsection{Stellar Metallicity and Planet Metallicity}
One may interpret present day stellar metallicity as a proxy for the metal-enrichment of the protoplanetary disk because the disk and star formed from the same molecular cloud. However, the mean metallicity of the protoplanetary disk may be different from the local disk metallicity at the location of planet formation. With this caveat in mind, we nonetheless looked for correlations between host star metallicity and the observed properties of sub-Saturns.

In contrast to equilibrium temperatures we observe a stronger set of correlations between the planetary properties of sub-Saturns and host star metallicity, \fe. Figure~\ref{fig:ss-metfe} shows \Mp, \fenv, \Mcore, and \Menv against \fe. Sub-Saturns orbiting metal-rich stars tend to be more massive. Interestingly, both the planetary core mass, \Mcore, and envelope mass, \Menv, appear to increase with stellar metallicity. This is understandable, however, given that planets with more massive cores should generally be able to accrete larger gaseous envelopes before their disks dissipate \citep{Lee15a}. This suggests that the metal-rich hosts had more solids available in their disk allowing those sub-Saturns to form more massive cores; and that these larger cores were then able accrete more massive envelopes. 

It is worthwhile to compare the observed \Mp--\fe trend to the stellar metallicity distribution of \Kepler planet hosts. \cite{buchhave:2012} observed that planets smaller than $\approx4~\Re$ are found around stars of wide-ranging metallicities ($-0.5 \lesssim$~\fe~$\lesssim+0.5$~dex) while larger planets are typically found around more metal-rich stars ($-0.2 \lesssim$~\fe~$\lesssim+0.5$~dex). This is consistent with our sample, constructed from planets found by \Kepler, \ktwo, and other surveys. Stellar metallicity of $> -0.2$~dex seems to be an important criterion for forming sub-Saturns. However, the additional metals in the disk seem to result in more massive final planets.

The path by which increased metallicity produces more massive sub-Saturns could proceed in one of two ways: (1) disks with more solid material form substantially more massive planet cores which grow smoothly into more massive planets or (2) disks with more solid material form slightly more massive cores which perturb neighboring planets causing collisions and mergers. We have a slight preference for the latter explanation, given the eccentricity distribution of sub-Saturns, explored in Section~\ref{ssec:eccentricity}. It is also plausible that metal-enriched disks could form planet cores more quickly, allowing for a long period of gas accretion. However, we see no evidence of this given the absence of a strong correlation between \fenv and host star metallicity.

In addition to the trends with stellar metallicity shown in Figure~\ref{fig:ss-metfe}, it is also interesting to examine whether there are trends in the heavy element abundances of sub-Saturns after controlling for the dependence on stellar metallicity. Recently, \citet{Thorngren16} examined planet metal mass fraction, $Z_P$, for 47 planets having $\Mp \approx 30-3000$~\Me relative to the heavy element fraction of their parent stars, $Z_\star = Z_{\odot}10^\fe$. $Z_P$ was computed via $Z_P = \Mcore/\Menv$ using thermal evolution models similar to those used here. \cite{Thorngren16} observed an anti-correlation between \Mp and $Z_P/Z_{\star}$. They argued that such an anti-correlation can be understood in terms of traditional core-accretion formation theory if one assumes that planets below the isolation mass are able accrete all of the solids in their isolation zone, typically $\approx$3.5 Hill-radii \citep{Lissauer93}, but not all their gas. Given these assumptions, Equation~9 of \citet{Thorngren16} predicts the planetary metal enrichment:
\begin{equation}
\label{eqn:metalratio}
\frac{Z_P}{Z_\star} = 3 f_H f_e \frac{H}{a}Q^{-1}\left(\frac{\Mp}{\Mstar}\right)^{-2/3}.
\end{equation}
Here, $f_H\sim3.5$ is the approximate number of Hill-radii from which a planet can effectively accrete solids, $f_e\sim1$ is an enrichment factor to allow for metal enhancement due to radial drift by solids, $H$ is the disk scale height, $a$ is the semi-major axis, and $Q$ is Toomre disk instability parameter \citep{Toomre64}. \citet{Thorngren16} found that Equation~\ref{eqn:metalratio} can reproduce the observed trend between \Mp and $Z_P/Z_\star$ if $f_H=3.5$, $f_e=1$, and $Q=5$.

\begin{figure*}
\plotone{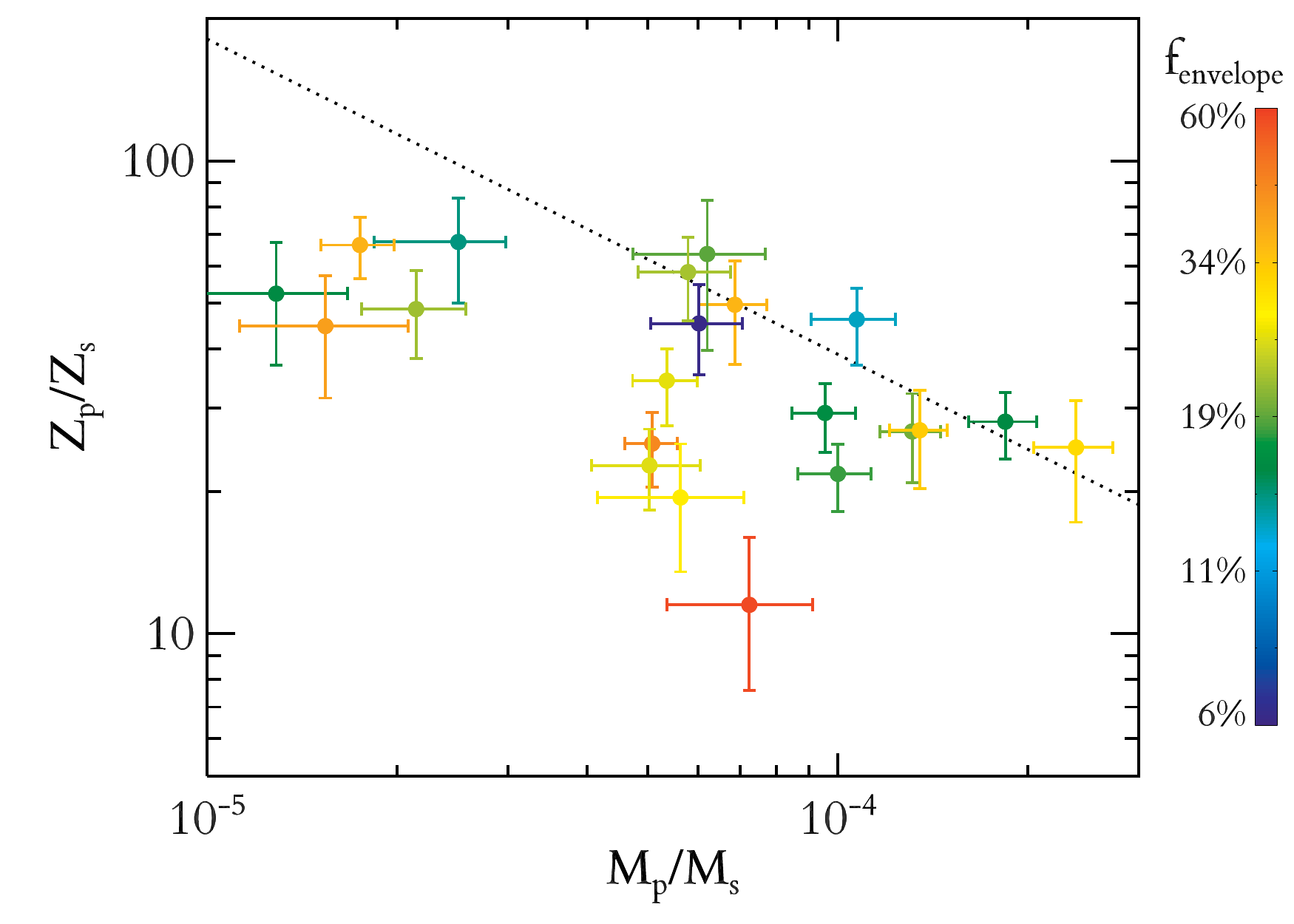}
\caption{The bulk heavy element enrichment of the sub-Saturns relative to their parent stars vs. the planet to star mass ratio. Planets are color-coded by their envelope mass fractions. The dotted line corresponds to the correlation found in \citet{Thorngren16} for massive planets $\Mp \approx 30-3000~\Me$.\label{fig:metalratio}}
\end{figure*}

Figure~\ref{fig:metalratio} compares our sample of sub-Saturns to the predictions of Equation~\ref{eqn:metalratio}. Sub-Saturns are a valuable laboratory for testing the physics of envelope accretion because they are larger than the sub-Neptunes, which typically have $\fenv \lesssim 10\%$, and the gas giants, which are nearly entirely envelope ($\fenv\sim100\%$). Following \cite{Thorngren16}, we approximate the planetary metal abundance by setting $Z_P = \Mcore/\Mp$. For the most massive Sub-Saturns (having $\Mp/\Mstar \gtrsim 10^{-4}$) where our sample overlaps with the \citet{Thorngren16} sample, we find good agreement with the predictions of Equation~\ref{eqn:metalratio}, as shown by the dotted line. Below $\Mp/\Mstar\sim10^{-4}$, however, we find a significant increase in dispersion below this relation, with many planets being significantly less enriched in metals than expected from Equation~\ref{eqn:metalratio}.

One interpretation is that this increase in scatter simply reflects the natural transition between giant planets, which essentially accrete all of the heavy elements in their feeding zones, and low mass planets which do not. Equation~\ref{eqn:metalratio}, assumes that the solids in the disk have fully decoupled from the gas, and that a planet can successfully accrete all of the solids near its isolation zone. At lower cores masses, gravitational focusing becomes more important, gas damping of planetesimals eccentricities becomes more efficient, and collision and growth timescales become longer. All of these factors mean less massive cores may not accrete all solids with in $\approx3.5$~Hill-radii, which may instead be dispersed or incorporated into other planets in compact, multi-planet systems. In summary, we examined whether the correlation between planet metal-enrichment and planet masses observed by \cite{Thorngren16} for planets having \Mp = 30--3000~\Me is present among the sub-Saturns. We do not observe a strong correlation, indicating a possible transition in the formation pathways of planets with $\Mp \lesssim 30$~\Me.

\subsection{Eccentricity and Planet Multiplicity}
\label{ssec:eccentricity}

The fits to K2-27, K2-39, and K2-108 RVs favored non-zero orbital eccentricities of $\val{K2-27-ecc}{e1}$, $\val{K2-39-ecc}{e1}$, and $\val{epic2117-ecc}{e1}$, respectively. Given that these planets are on short orbital periods of 4.6--6.8~d, respectively, it is worthwhile to consider the extent to which tides are expected to damp away eccentricity. The timescale for eccentricity damping \citep{Goldreich66} is given by 
\begin{equation}
\label{eqn:taue}
    \tau_e = \frac{4}{63} 
                 \left(\frac{\Qprime}{n}\right) 
                 \left(\frac{\Mp}{\Mstar}\right)
                 \left(\frac{a}{\Rp}\right)^5.
\end{equation}
Here, $n = \sqrt{G \Mstar / a^3}$ is the mean motion, and \Qprime, the modified tidal quality factor, is given by $\Qprime = 3 Q / 2 k_2$, where $Q$ is the specific dissipation function and $k_2$ is the Love number (see \citealt{Goldreich66,Murray99,Mardling04}). \Qprime is quite uncertain even for planets in the Solar System. As a point of reference \cite{Lainey15} give $\Qprime \approx 6,000-18,000$ for Saturn, based on Cassini ranging data. \cite{Tittemore89,Tittemore90} give $Q \approx 11,000-39,000$ for Uranus based on studies of the Uranian satilites, which translates to $\Qprime \approx 165,000-585,000$, adopting $k_2 = 0.104$ from \citep{Gavrilov77}. \Qprime is even more uncertain for sub-Saturns which have no Solar System analogs. Here, we adopt $\Qprime = 10^5$ for the sub-Saturns, with the understanding that this estimate is only good to order of magnitude. Re-writing Equation~\ref{eqn:taue},

\begin{eqnarray}
    \tau_e & \sim 1.8~\mathrm{Gyr} 
                \left(\frac{\Qprime}{10^5}\right) 
                \left(\frac{P}{10~\mathrm{d}}\right)
                \left(\frac{\Mp}{10~\Me}\right)
                \left(\frac{\Mstar}{\Msun}\right)^{-1} \nonumber\\ 
           &    \times \left(\frac{a}{0.05~\AU}\right)^5
                \left(\frac{\Rp}{4 \Re}\right)^{-5}.  \nonumber 
\end{eqnarray}

For K2-27b, we find $\tau_e \sim 10$~Gyr, comparable to the age of the system, suggesting that if this eccentricity was caused by planet-planet scattering early in the star's lifetime, the eccentricity could persist to the present day. K2-39b and K2-108b are slightly larger than K2-27b and also orbit closer to their host stars. Because the circularization timescale is a strong function of $a/\Rp$, they have substantially shorter $\tau_e$ of $\sim 0.6$~Gyr and $\sim2$~Gyr, respectively. These circularization timescales are formally shorter than the age of their host stars and present some challenges for understanding any present day eccentricities. This tension could be resolved if \Qprime is $\sim10^6$ as opposed to the assumed value of $\sim10^5$. Eccentric orbits could also be maintained by additional, yet undetected planets. \cite{Deming07} proposed such an explanation for GJ436b, another short-period eccentric sub-Saturn (see Table~\ref{tab:subsaturns}). Assuming $\Qprime\sim10^5$, $\tau_e$ for GJ436b is only $\sim 0.1$~Gyr. While no additional planets have been detected in the GJ436 system, \cite{Batygin09} showed this explanation to be plausible in the context of secular theory.

\begin{figure*}
\plotone{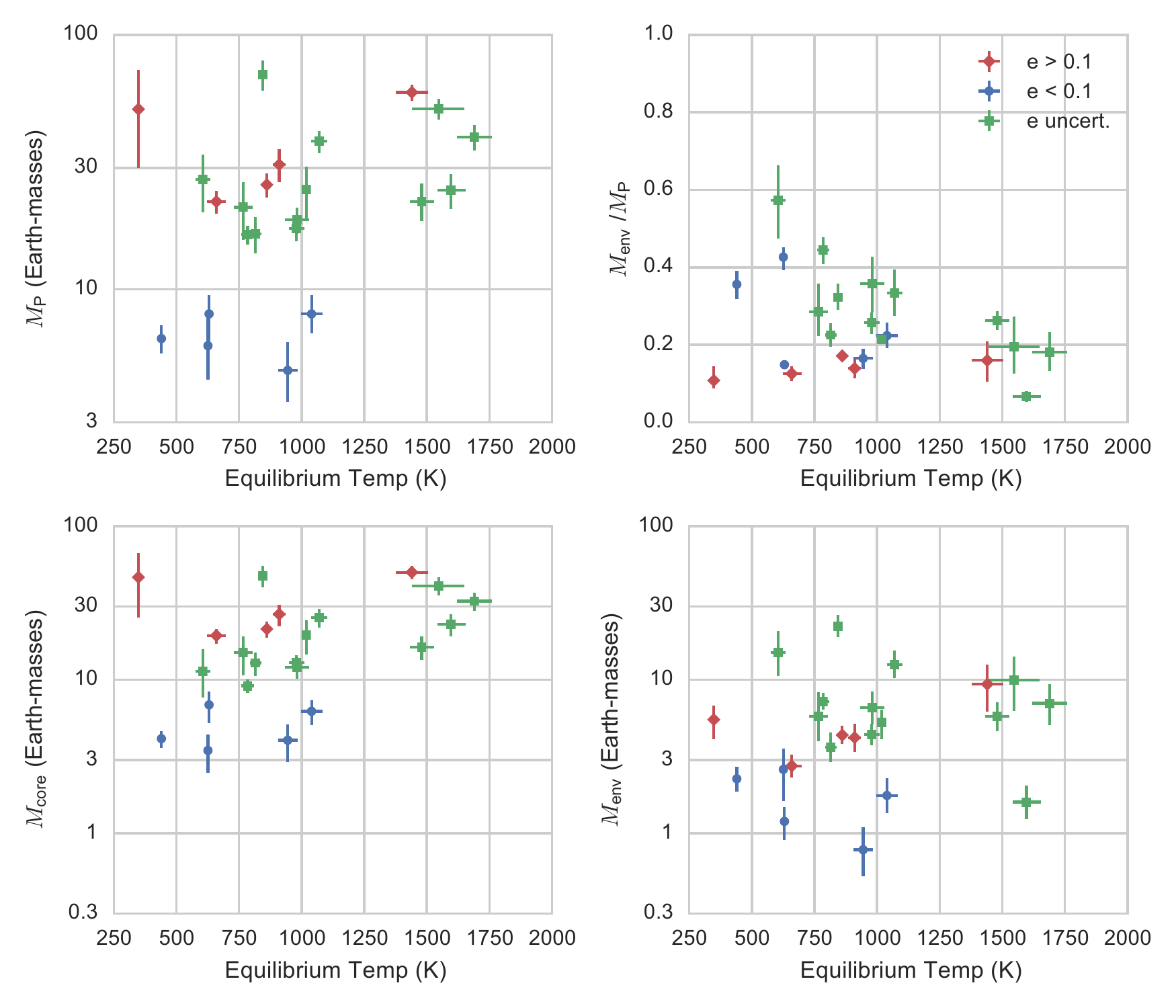}
\caption{Panels a--d show the planet mass (\Mp), envelope fraction (\Mfenv), core mass (\Mcore), and envelope mass (\Menv) as a function of planet blackbody equilibrium temperature (\Teq). We observe an absence of planets having high envelope fractions ($\fenv \gtrsim 0.25$) at high equilibrium temperatures as expected from photo-evaporation. However, the lack of a strong trend between \Teq and \fenv suggests that photo-evaporation has not significantly sculpted the majority of sub-Saturns shown. Planets where eccentricity has been constrained to be less than or greater than 0.1 are colored blue and red, respectively.\label{fig:teq}}
\label{fig:}
\end{figure*}

\begin{figure*}
\plotone{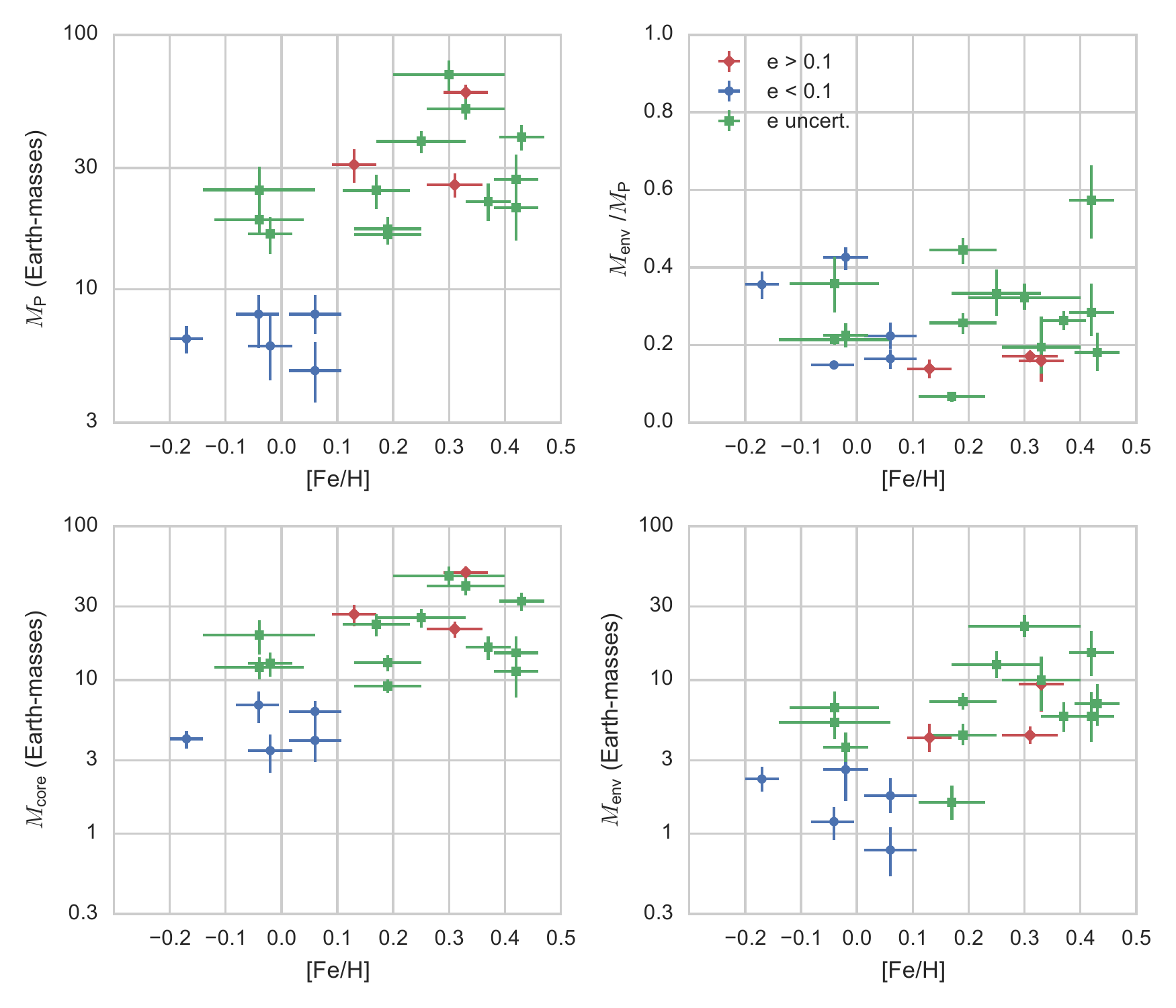}
\caption{Same as Figure~\ref{fig:pnum} but showing planet mass (\Mp), envelope fraction (\Mfenv), core mass (\Mcore), and envelope mass (\Menv) as a function of stellar metallicity, \fe. We observe a correlation between metal-rich host stars and more massive sub-Saturns. The host star metallicity does not correlate with \fenv, suggesting that disk metallicity is not the only factor that affects the final add-mixture of envelope and solids that comprise these planets.\label{fig:fe}}
\label{fig:ss-metfe}
\end{figure*}

\begin{figure*}
\plotone{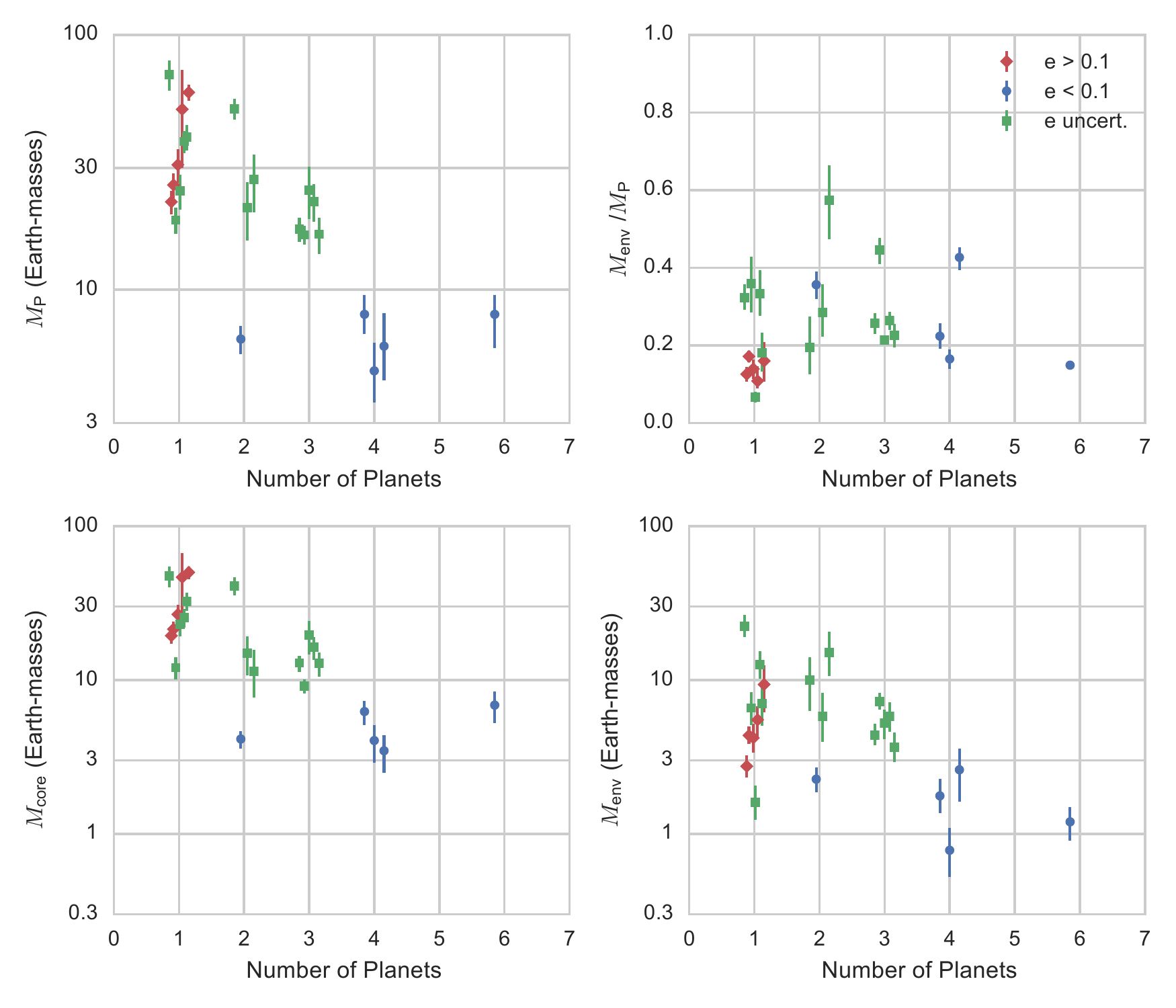}
\caption{Panels a--d show the planet mass (\Mp), core mass fraction (\Mcmf), core mass (\Mcore), and envelope mass (\Menv) as a function of the number of detected planets in the system. The X-coordinates of each planet has been offset for legibility. The most massive sub-Saturns tend reside in low-multiplicity systems.\label{fig:pnum}}
\end{figure*}

In Table~\ref{tab:subsaturns}, we also included the measured eccentricities, when available.  We broke the sample into low ($e$ < 0.1), moderate ($e$ > 0.1), and poorly-constrained eccentricities. In order for a planet to be included in the low/moderate eccentricity bins, its entire $1\sigma$ eccentricity confidence interval must be below/above 0.1. Planets with eccentricity upper limits or constraints straddling 0.1 are fall in the poorly-constrained category. These different eccentricity categories are color-coded in Figures~\ref{fig:teq}, \ref{fig:ss-metfe}, and \ref{fig:pnum}.

We observed that the highest-mass planets were often the only detected planet in the system. Figure~\ref{fig:pnum} shows \Mp, \fenv, \Mcore, and \Menv as a function of the total number of detected planets in the system. There is a steady decline in the mass of sub-Saturns as overall multiplicity increases. The planets with moderate eccentricities are confined to apparently single systems. These high-mass singles could have originally had neighbors, but were in a dynamically unstable architecture. 

Such instabilities would eventually lead to close encounters and planet-planet scattering. Because these planets are so deep in the potential wells of their host stars, these scattering events would likely lead to mergers as opposed to ejections from the system. The maximum velocity a planet can impart to its neighbor is the escape velocity, 
\[
\vesc = 11.2~\kms \left(\frac{\Mp}{\Me}\right)^{1/2} \left(\frac{\Rp}{\Re}\right)^{-1/2}.
\]
and if \vesc is smaller than the orbital velocity,
\[
\vorb = 30~\kms \left(\frac{\Mstar}{\Msun}\right)^{1/2} \left(\frac{a}{1~\AU}\right)^{-1/2},
\]
single scattering events cannot lead to ejections. For the sub-Saturns in Table~\ref{tab:subsaturns}, $\vesc/\vorb \approx 0.1-0.3$.%
\footnote{
We have excluded Kepler-413 b, because it is a circumbinary planet with more complex criteria for ejection.
}
Any previous dynamical instabilities would likely lead to planet mergers, increasing their total mass. The present-day eccentricities of the more massive sub-Saturns may be a relic of previous scattering and merging events. 

It is worth considering the biases associated with the different techniques by which planet mass and eccentricity are measured and whether they could be responsible for the observed mass-multiplicity-eccentricity trends. TTV measurements require multi-planet system and thus do not contribute to any points in the $N_P = 1$ bin in Figure~\ref{fig:pnum}. Constraining $e$ < 0.1 is challenging with RVs given that one is trying to measure slight deviations from sinusoidal RV curves. Therefore, limitations of the RV and TTV techniques might explain why there are no single planets with secure eccentricity measurements of < 0.1.

However, these observational biases cannot explain why planets in multi-planet systems are low mass and preferentially circular. Previous studies (e.g. \citealt{Weiss14}) have noted that planets with TTV mass measurements are typically less massive than planets with RV mass measurements. TTVs, however, are not blind to high-mass planets; such planets would produce {\em larger} TTVs. The lack of high-mass, high-eccentricity planets in multi-planet systems is likely the result of dynamical instabilities. Planets in such systems would likely perturb one another and merge, resulting high-mass planets in low-multiplicity systems.


\begin{deluxetable*}{lRRRRRRRR}
\tablecaption{Sub-Saturns with well-measured densities\label{tab:subsaturns}}
\tablecolumns{9}
\tablehead{
        \colhead{Name} & 
        \colhead{$N_P$} & 
        \colhead{\Rp} & 
        \colhead{\Mp} & 
        \colhead{$\rho$} & 
        \colhead{$e$} & 
        \colhead{\Teq} & 
        \colhead{\fenv} & 
        \colhead{\fe} \\
        \colhead{} & 
        \colhead{} & 
        \colhead{\Re} & 
        \colhead{\Me} & 
        \colhead{\gcc} & 
        \colhead{} & 
        \colhead{K} & 
        \colhead{\%} & 
        \colhead{dex} 
        }
\tablewidth{0pt}
\startdata
Kepler-4 b & 1 & 4.00^{ +0.21 }_{ -0.21 } & 24.5^{ +3.8 }_{ -3.8 } & 2.09^{ +0.53 }_{ -0.44 } & \nodata  & 1597 & 6.7^{ +1.3 }_{ -1.4 } & +0.17\\
GJ 436 b & 1 & 4.17^{ +0.17 }_{ -0.17 } & 22.1^{ +2.3 }_{ -2.3 } & 1.67^{ +0.30 }_{ -0.26 } & 0.1383^{ +0.0002 }_{ -0.0002 } & 659 & 12.6^{ +1.9 }_{ -1.9 } & \nodata \\
Kepler-11 e & 6 & 4.19^{ +0.07 }_{ -0.09 } & 8.0^{ +1.5 }_{ -2.1 } & 0.60^{ +0.15 }_{ -0.14 } & 0.0120^{ +0.0060 }_{ -0.0060 } & 630 & 14.9^{ +0.7 }_{ -0.7 } & -0.04\\
Kepler-413 b & 1 & 4.35^{ +0.10 }_{ -0.10 } & 51.0^{ +22.0 }_{ -21.0 } & 2.40^{ +1.00 }_{ -1.00 } & 0.1185^{ +0.0018 }_{ -0.0017 } & 348 & 10.8^{ +3.8 }_{ -2.0 } & \nodata \\
K2-27 b & 1 & 4.48^{ +0.23 }_{ -0.23 } & 30.9^{ +4.6 }_{ -4.6 } & 1.88^{ +0.46 }_{ -0.38 } & 0.2510^{ +0.0880 }_{ -0.0880 } & 910 & 13.9^{ +2.4 }_{ -2.5 } & +0.13\\
Kepler-223 e & 4 & 4.60^{ +0.27 }_{ -0.41 } & 4.8^{ +1.4 }_{ -1.2 } & 0.27^{ +0.11 }_{ -0.09 } & 0.0510^{ +0.0190 }_{ -0.0190 } & 944 & 16.5^{ +2.5 }_{ -2.6 } & +0.06\\
HAT-P-11 b & 1 & 4.73^{ +0.16 }_{ -0.16 } & 25.7^{ +2.9 }_{ -2.9 } & 1.34^{ +0.22 }_{ -0.20 } & 0.1980^{ +0.0460 }_{ -0.0460 } & 861 & 17.1^{ +1.5 }_{ -1.5 } & +0.31\\
K2-32 b & 3 & 5.13^{ +0.28 }_{ -0.28 } & 16.5^{ +2.7 }_{ -2.7 } & 0.67^{ +0.18 }_{ -0.15 } & \nodata  & 815 & 22.5^{ +3.0 }_{ -3.1 } & -0.02\\
Kepler-25 c & 3 & 5.20^{ +0.09 }_{ -0.09 } & 24.6^{ +5.7 }_{ -5.7 } & 0.96^{ +0.24 }_{ -0.23 } & \nodata  & 1018 & 21.4^{ +1.3 }_{ -1.3 } & -0.04\\
Kepler-223 d & 4 & 5.24^{ +0.26 }_{ -0.45 } & 8.0^{ +1.5 }_{ -1.3 } & 0.30^{ +0.10 }_{ -0.07 } & 0.0370^{ +0.0180 }_{ -0.0170 } & 1040 & 22.3^{ +3.4 }_{ -3.2 } & +0.06\\
K2-108 b & 1 & 5.28^{ +0.54 }_{ -0.54 } & 59.4^{ +4.4 }_{ -4.4 } & 2.21^{ +0.90 }_{ -0.59 } & 0.1800^{ +0.0420 }_{ -0.0420 } & 1440 & 16.0^{ +4.8 }_{ -5.4 } & +0.33\\
Kepler-18 c & 3 & 5.49^{ +0.26 }_{ -0.26 } & 17.3^{ +1.9 }_{ -1.9 } & 0.57^{ +0.12 }_{ -0.10 } & \nodata  & 979 & 25.7^{ +2.6 }_{ -2.8 } & +0.19\\
K2-24 b & 2 & 5.68^{ +0.56 }_{ -0.56 } & 21.0^{ +5.4 }_{ -5.4 } & 0.62^{ +0.31 }_{ -0.22 } & \nodata  & 766 & 28.4^{ +7.4 }_{ -6.2 } & +0.42\\
K2-39 b & 1 & 5.71^{ +0.63 }_{ -0.63 } & 39.7^{ +4.6 }_{ -4.6 } & 1.16^{ +0.54 }_{ -0.34 } & 0.1500^{ +0.0760 }_{ -0.0760 } & 1689 & 18.1^{ +5.2 }_{ -4.8 } & +0.43\\
Kepler-101 b & 2 & 5.77^{ +0.85 }_{ -0.79 } & 51.1^{ +5.1 }_{ -4.7 } & 1.46^{ +0.90 }_{ -0.50 } & 0.0860^{ +0.0800 }_{ -0.0590 } & 1547 & 19.5^{ +7.9 }_{ -6.9 } & +0.33\\
Kepler-87 c & 2 & 6.14^{ +0.29 }_{ -0.29 } & 6.4^{ +0.8 }_{ -0.8 } & 0.15^{ +0.03 }_{ -0.03 } & 0.0390^{ +0.0120 }_{ -0.0120 } & 440 & 35.6^{ +3.4 }_{ -3.7 } & -0.17\\
HATS-7 b & 1 & 6.31^{ +0.52 }_{ -0.38 } & 38.1^{ +3.8 }_{ -3.8 } & 0.83^{ +0.23 }_{ -0.17 } & \nodata  & 1070 & 33.3^{ +6.1 }_{ -5.8 } & +0.25\\
HAT-P-26 b & 1 & 6.33^{ +0.81 }_{ -0.36 } & 18.8^{ +2.2 }_{ -2.2 } & 0.40^{ +0.15 }_{ -0.10 } & 0.1240^{ +0.0600 }_{ -0.0600 } & 981 & 35.8^{ +6.9 }_{ -7.5 } & -0.04\\
CoRoT-8 b & 1 & 6.39^{ +0.22 }_{ -0.22 } & 69.9^{ +9.5 }_{ -9.5 } & 1.47^{ +0.27 }_{ -0.25 } & \nodata  & 844 & 32.2^{ +3.5 }_{ -3.1 } & +0.30\\
Kepler-56 b & 3 & 6.51^{ +0.29 }_{ -0.28 } & 22.1^{ +3.9 }_{ -3.6 } & 0.44^{ +0.10 }_{ -0.09 } & \nodata  & 1479 & 26.3^{ +2.4 }_{ -2.4 } & +0.37\\
Kepler-18 d & 3 & 6.98^{ +0.33 }_{ -0.33 } & 16.4^{ +1.4 }_{ -1.4 } & 0.26^{ +0.05 }_{ -0.04 } & \nodata  & 784 & 44.5^{ +3.2 }_{ -3.6 } & +0.19\\
Kepler-79 d & 4 & 7.16^{ +0.13 }_{ -0.16 } & 6.0^{ +2.1 }_{ -1.6 } & 0.09^{ +0.03 }_{ -0.03 } & 0.0250^{ +0.0590 }_{ -0.0230 } & 626 & 42.6^{ +2.5 }_{ -3.3 } & -0.02\\
K2-24 c & 2 & 7.82^{ +0.72 }_{ -0.72 } & 27.0^{ +6.9 }_{ -6.9 } & 0.31^{ +0.14 }_{ -0.10 } & \nodata  & 605 & 57.3^{ +9.1 }_{ -9.9 } & +0.42\\

\enddata
\tablecomments{List of sub-Saturns having densities measured to 50\% or better, assembled from the NASA Exoplanet Archive, this work, and other sources. $N_P$ ---Total number of detected planets in the system, $e$---orbital eccentricity, we do not report eccentricity if only upper limits are available. \Teq refers to the blackbody temperature (i.e. assuming zero albedo). \fenv---``envelope fraction'' fraction of planet's mass in H/He in the two-component modeling of planet mass and radius, described in Section~\ref{ssec:cmf}. Notes on individual systems: Kepler-4 b---\cite{Borucki10b}; GJ 436 b---\cite{Butler04}; Kepler-11 e---\cite{Lissauer13}; Kepler-413 b---\cite{Kostov14}; K2-27 b---This work; Kepler-223 e---\cite{Mills16}; HAT-P-11 b---\cite{Bakos10}; K2-32 b---This work; Kepler-25 c---\cite{Marcy14}; Kepler-223 d---\cite{Mills16}; K2-108b---This work; Kepler-18 c---\cite{Cochran11}; K2-24 b---\cite{Petigura16}; K2-39 b---This work; Kepler-101 b---\cite{Bonomo14}; Kepler-87 c---\cite{Ofir14}; HATS-7 b---\cite{Bakos15}; HAT-P-26 b---\cite{Hartman11}; CoRoT-8 b---\cite{Borde10}, adopted $3\pm1$~Gyr; Kepler-56 b---\cite{Huber13a}; Kepler-18 d---\cite{Cochran11}; Kepler-79 d---\cite{Jontof-Hutter14}; K2-24 c---\cite{Petigura16}.}
\end{deluxetable*}

\section{Conclusions}
\label{sec:conclusions}

We presented radial velocity measurements of four systems hosting sub-Saturn planets observed by the \ktwo mission: K2-27, K2-32, K2-39, and K2-108. These RVs enabled mass measurements of 16\% or better and detailed analysis of the planetary heavy element fraction. Despite the similar sizes of the planets, their masses range from 16--60~\Me, implying widely different core and envelope masses. Despite the differences in the masses of these planets, the fraction of their mass in H/He is similar $\approx$80\%. This trend is seen in the population of $\approx$20 sub-Saturns with well-measured masses. 

Sub-Saturns as a class of planets show a remarkable diversity in mass \Mp = 6--60 \Me, with little dependence on planet size. We observe a strong correlation between stellar metallicity and planet mass. This implies that metal-rich disks produce more, or more massive, cores. Finally, we observe a tendency of the most massive sub-Saturns to have moderate eccentricities and to reside in apparently single systems. Future observational and theoretical work will further illuminate  these mysterious planets, absent in our own Solar System.

\acknowledgments 
We thank Konstantin Batygin and John Livingston for helpful discussions.  E.~A.~P.\ acknowledges support from a Hubble Fellowship grant HST-HF2-51365.001-A awarded by the Space Telescope Science Institute, which is operated by the Association of Universities for Research in Astronomy, Inc. for NASA under contract NAS 5-26555. This work has made use of data from the European Space Agency (ESA) mission \Gaia, processed by \Gaia Data Processing and Analysis Consortium (DPAC). Funding for the DPAC has been provided by national institutions, in particular the institutions participating in the \Gaia Multilateral Agreement. Some of the data presented herein were obtained at the W.~M.~Keck Observatory (which is operated as a scientific partnership among Caltech, UC, and NASA). We thank the Caltech and NASA Keck Time Allocation Committees for providing HIRES time. This work included observations obtained at the Gemini Observatory, which is operated by the Association of Universities for Research in Astronomy, Inc., under a cooperative agreement with the NSF on behalf of the Gemini partnership: the National Science Foundation (United States), the National Research Council (Canada), CONICYT (Chile), Ministerio de Ciencia, Tecnolog\'{i}a e Innovaci\'{o}n Productiva (Argentina), and Minist\'{e}rio da Ci\^{e}ncia, Tecnologia e Inova\c{c}\~{a}o (Brazil).The authors wish to recognize and acknowledge the very significant cultural role and reverence that the summit of Maunakea has always had within the indigenous Hawaiian community.  We are most fortunate to have the opportunity to conduct observations from this mountain.

\software{Numpy/Scipy \citep{VanDerWalt11}, Matplotlib \citep{Hunter07}, Pandas \citep{McKinney10}, Astropy \citep{Astropy13}, emcee \citep{Goodman10,Foreman-Mackey13}, SME \citep{Brewer15}, isochrones \citep{Morton15}, k2sc \citep{Aigrain15}, batman \citep{Kreidberg15}, radvel (https://github.com/California-Planet-Search/radvel),  k2phot (https://github.com/petigura/k2phot)}

\bibliography{revtex-custom,manuscript}
\end{document}